\documentclass[journal,comsoc]{IEEEtran}

\usepackage[T1]{fontenc}
\usepackage{graphicx}
\usepackage{subfigure} 
\usepackage{caption}
\usepackage{mathrsfs}
\usepackage{amsmath,amssymb,amsfonts}
\usepackage{url}
\usepackage{algorithmic}
\usepackage[linesnumbered,ruled,vlined]{algorithm2e}
\usepackage{multicol}
\usepackage{setspace}

\ifCLASSINFOpdf
\else

\fi
\usepackage{amsmath}
\begin{document}
\begin{spacing}{0.97}
\title{Secure and Efficient Blockchain based Knowledge Sharing for Intelligent Connected Vehicles}

\author{Haoye~Chai,
        Supeng~Leng*,
        Fan~Wu,
        and~Jianhua~He
\thanks{H. Chai, S. Leng, F. Wu are with the School of Information and Communication Engineering, University of Electronic Science and Technology of China (UESTC), Chengdu 611731, China, and Shenzhen Institute for Advanced Study, UESTC.\\
J. He is with the School of Computer Science and Electronic Engineering, University of Essex, Colchester, CO4 3SQ, UK (e-mail: j.he@essex.ac.uk).\\
* Corresponding author, email: spleng@uestc.edu.cn.}}

\markboth{Journal of \LaTeX\ Class Files,~Vol.~14, No.~8, August~2015}%
{Shell \MakeLowercase{\textit{et al.}}: Bare Demo of IEEEtran.cls for IEEE Communications Society Journals}
\maketitle

\begin{abstract}
The emergence of Intelligent Connected Vehicles (ICVs) shows great potential for future intelligent traffic systems, enhancing both traffic safety and road efficiency. 
However, the ICVs relying on data driven perception and driving models face many challenges, including the lack of comprehensive knowledge to deal with complicated driving context. 
In this paper, we are motivated to investigate cooperative knowledge sharing for ICVs. 
We propose a secure and efficient directed acyclic graph (DAG) blockchain based knowledge sharing framework, 
aiming to cater for the micro-transaction based vehicular networks.
The framework can realize both local and cross-regional knowledge sharing.
Then, the framework is applied to autonomous driving applications, wherein machine learning based models for autonomous driving control can be shared. 
A lightweight tip selection algorithm (TSA) is proposed for the DAG based knowledge sharing framework to achieve consensus and identity verification for cross-regional vehicles. 
To enhance model accuracy as well as minimizing bandwidth consumption,
an adaptive asynchronous distributed learning (ADL) based scheme is proposed for model uploading and downloading.
Experiment results show that the blockchain based knowledge sharing is secure, and it can resist attacks from malicious users.
In addition, the proposed adaptive ADL scheme can enhance driving safety related performance compared to several existing algorithms. 
\end{abstract}

\begin{IEEEkeywords}
Knowledge Sharing, DAG blockchain, Intelligent Connected Vehicles, Distributed Learning
\end{IEEEkeywords}

\IEEEpeerreviewmaketitle

\section{Introduction}

\IEEEPARstart{W}{ith} the development of vehicular communications and artificial intelligence, intelligent connected Vehicles (ICVs) is receiving increasing interest as a promising technology for tackling the challenges faced by the intelligent transportation systems \cite{ICV1}, \cite{ICV2}. 
intelligent connected vehicles (ICVs) is the combination of Connected Vehicles (CVs) and Autonomous Vehicles (AVs). Served by the ubiquitous Vehicle to Everything (V2X) technologies, CVs is capable of sharing sensing data and driving information with other vehicles and traffic infrastructures to improve the traffic safety and efficiency. Meanwhile, equipped with sensors and intelligent on-board modules, AVs can support intelligent vehicular applications and various levels of driving automation, such as Advanced Driving Assistance System (ADAS) and self-driving. Several advanced driving use cases (including extended sensors and cooperative driving) have been specified in the fifth generation (5G) standards. By integrating the advantages of CVs and AVs, the ICVs is an important component of future Intelligent Transportation Systems (ITS).
However, the perception and driving models behind the ICVs are mainly data driven (e.g., trained by huge sensing and driving data). They are lack of knowledge to deal with unseen and complex driving scenarios,
and security and privacy needs to be protected in dynamic vehicular network environment. 
To address these problems, we are motivated to extend the sharing of raw sensing data to the sharing of knowledge for ICVs. 

Knowledge sharing is defined as the activities of transferring or disseminating knowledge among a group of people or organizations \cite{no1}. 
Semantic webs and knowledge graphs have been widely used for knowledge sharing in many applications, 
such as Question and Answer (Q\&A) systems, recommendation and webpage searching systems. 
With the aid of emerging Distributed Learning (DL) technology, 
knowledge sharing for ICVs can provide great benefits to enhance road safety and driving experiences. ICVs can train their own sensing or driving data by the on-board computing devices, and obtain corresponding parameters ($e.g.$ models/policies) as refined knowledge.
Various levels of knowledge (such as traffic statistics, traffic control, driving rules, sensing and driving models, and crowd sourced maps) 
can be shared among the ICVs and the roadside infrastructures. In this case, DL based knowledge sharing enhances the perception and comprehension of the driving environment, and support driving decision making \cite{xiong}. 

Despite the great potentials held by knowledge sharing of ICVs,  several critical issues remains to be addressed. 
On the one hand, in the DL based knowledge sharing process, attackers could manipulate the shared knowledge or spread misleading knowledge. 
While central entities such as roadside units (RSUs) could be employed to manage knowledge sharing, they are subjected to various attacks such as single point of failure. 
On the other hand,  existing DL based knowledge sharing shows weakness in sharing efficiency. There is a lack of efficient  management and cooperation of the distributed knowledge that shared by ICVs.  
For example, conventional federated learning enforce workers to train and share their models in a synchronous manner, resulting in a long delay due to the intermittent links among ICVs.

To address the above problems, we resort to the emerging DAG blockchain and asynchronous distributed learning (ADL) technologies, aiming to achieve secure and efficient knowledge sharing in highly dynamic vehicular networks.
Compared to the existing blockchains such as Bitcoin and Ethereum, utilizing block mining process with Proof-of-work and Proof-of-Stake that can incur huge computation and communication loads,
there is no concept of blocks in DAG, and the fundamental unit of DAG is called site that includes a micro-transaction.  The new site can be directly appended on the DAG without the mining process, thus realizing the lightweight and resource-saving. Besides, the consensus of DAG is based on a tip selection algorithm (TSA), which utilize the subsequent new sites to verify old sites.  The more sites are issued, the faster verifications are implemented, thus it is highly desirable for micro-transaction based mobile and distributed vehicular networks.  

Built on the top of secure context by DAG, the ADL shows great potential for enhancing the correlation of shared knowledge cached in the DAG ledger.  By utilizing particular updating schemes, the ADL servers aggregate the distributed local knowledge and obtain a comprehensive model to enhance the accuracy of knowledge.
Compared with conventional synchronous distributed learning \cite{FedAve}, the ADL supports ICVs to upload their models in an asynchronous way,  
and the ICVs can obtain the on-the-fly updated model without waiting for others, which is suitable for the high dynamic vehicular networks.

Therefore, we propose a directed acyclic graph (DAG) blockchain based knowledge sharing framework. The framework exploits a consortium blockchain architecture, wherein ICVs share their trained local models as knowledge, and issue the models as sites to the DAG blockchain. RSUs act as the DAG nodes to maintain the ledger, as well as being the tamper-proof, attack-resistance ADL servers to aggregate the distributed knowledge from the DAG ledger.
It is noted the framework is general for different ICV applications, thereby different DAG and ADL related schemes can be designed to cater for the quality of service (QoS) of different applications.
In order to show the potential of the proposed framework, a specific application to autonomous driving is investigated,  wherein driving control related models can be shared by ICVs. 
The contributions of the paper can be summarized as follows:
\begin{itemize}
\item We propose a secure and efficient knowledge sharing framework for ICVs based on DAG blockchain, wherein both local and cross-regional knowledge sharing are integrated.  
The DAG blockchain can guarantee the security and tamper-proof of the shared knowledge that is important for ICV safety and driving applications. 
The RSUs act as both DAG nodes and ADL servers, so that it can aggregate and manage the distributed knowledge from the DAG ledger to enhance the knowledge sharing efficiency.

\item Built on the top of the framework,  we develop a DAG blockchain for autonomous driving application.
A reversed two hop tip selection algorithm is proposed to achieve lightweight consensus that takes driving style of ICVs into consideration during tip approval process. 
A classified DAG ledger is designed to record the shared knowledge of ICVs, aiming to enhance the correlation of the knowledge. 
Based on the proposed algorithm, an identity verification scheme for cross-region ICVs is proposed without introducing additional storage costs. 
The convergence of the ledgers in the DAG blockchain system is proved with mathematic analysis.

\item In order to minimize bandwidth consumption as well as improving model accuracy, an adaptive asynchronous distributed learning scheme is proposed for the uploading and aggregating processes of shared knowledge.
The convergence of the proposed ADL is proved, and the optimal model weight and ICV driving style are analyzed based on the optimal convergence point.
%
\end{itemize}

The remainder of this paper is organized as follows. We present related works of knowledge sharing, blockchain and distributed learning in Section II. The overview of the proposed DAG based knowledge sharing framework is presented in Section III. In Section IV, the blockchain knowledge sharing is applied for autonomous driving application.  The system components are presented and the convergence of the blockchain ledger is proved. 
The design and convergence analysis of the ADL algorithm is presented in Section V.  In Section VI, experiment and simulation results for the blockchain knowledge sharing and ADL algorithm are presented and discussed. Finally, the paper is concluded in Section VII.

\section{Related Work}

Knowledge sharing in internet of vehicles (IoV) has been studied recently. By utilizing the information of vehicular trajectories, a clustering scheme was proposed based on the semi-Markov process \cite{clustering}. Rula $et.al$ proposed a data fusion algorithm to realize vehicular context awareness \cite{context_awareness}. A joint rate control and resource allocation scheme was developed with the information of channel states and delay constraints \cite{joint}. During the  information sharing process, it is crucial to guarantee the security and reliability, which has not been addressed in these works. Tampered or malicious knowledge can cause severe safety issues and degrade system performance. In this case, the emerging blockchain technology shows strong resistance to malicious attacks, and it has attracted increasing attention.

In literature \cite{bc_know1}, a user-centric blockchain (UCB) framework was proposed to preserve the reliability of edge data, in which a lightweight consensus mechanism was developed. To maintain the traceability of data, Lin $et \ al.$ presented a consortium blockchain \cite{bc_know2}. For ICVs, a service-oriented public blockchain was proposed for ride-sharing that aims to guarantee trust and fair payment during the sharing process \cite{bc_know4}. Cui $et \ al.$ proposed a blockchain-enabled payment paradigm in which the data is encrypted and outsourced decryption is payable \cite{bc_know5}. However, most existing works utilized $"block"$ data structures to store data in blockchains, which introduces additional packing expenditure and mining consumption. Moreover, the $"chain"$ structure leads to a time-consuming consensus process and a slow data recording, which is not suitable for large-scale ICV networks. Therefore, the existing blockchain frameworks are not readily applicable to the ICV network.

To further enhance the efficiency of knowledge sharing,  DL has been studied for vehicular networks.  A DL framework was developed to maintain the communication efficiency of IoV, and packet loss probability and throughput were analyzed under the framework \cite{fl_iov6}. Zhang $et \ al.$ proposed a two time scales DL algorithm to implement mode selection and resource allocation for vehicular networks \cite{fl_iov7}.  
Although the above studies have made great efforts on knowledge sharing, they are either insecure or not efficient for dynamic ICV networks. And they have not consider the ICV driving safety applications.

\section{ Design of the Knowledge Sharing Framework}
We aim to design an efficient and secure knowledge sharing framework for ICVs, in which ICVs extract knowledge from raw data and share the knowledge with others.
In order to ensure the security of knowledge sharing, we propose a DAG blockchain-based approach to encapsulate sharing knowledge.
The knowledge sharing framework is shown in Fig. 1. The whole ICV network is divided into multiple traffic regions. In each region, there is a consortium-based DAG blockchain,  where RSUs act as DAG nodes to cache the DAG ledger. Vehicles within one region will continuously collect data and extract the knowledge from environment. They share the knowledge as $site$s that are sent to the RSUs for consensus. 

The basic workflow of the knowledge sharing process can be described by the following four steps.

(1). Knowledge Extracting and Site Generation: By utilizing sensors and computing devices,  ICVs  extract valuable knowledge $K^{s}$ from the original raw data, such as traffic flow trend and driving configuration parameters.  The term $s$ is the impact scope of knowledge, which is regarded as the criterion for the judgement of cross-region knowledge sharing.
After completing a simple proof-of-work calculation $c$, the vehicle encapsulates the shared knowledge $K^s$ together with $c$ into a site. The site includes the signature of the vehicle $Sig$ that is generated by its private key, and the hash value $H$ of the site. 
The typical format of a site $S_v$ is as follows,
\begin{equation}
S=\{H, K^s_v, f,  c_v, \textbf{w}, Sig_v\}
\end{equation}
where $\textbf{w}$ is the vector of its own weight and cumulative weight of one site.  $f$ is the feature term of the shared knowledge, which contains different features according to different applications.

(2). Knowledge Verifying and Appending: The encapsulated site is then broadcast to vehicular network and append to the DAG ledger. 
The issuing ICV will choose and verify two existing tips on the DAG ledger according to the tip selection algorithm. The verification method varies depending on the application scenario.

(3). Local Knowledge Sharing: 
Within each traffic region,  a consortium based DAG architecture is adopted.
RSUs act as the blockchain nodes to cache and synchronize the local DAG ledger. 
At the same time, the RSUs can aggregate the knowledge that is cached in the DAG ledger, and act as the ADL servers to update the distributed knowledge to obtain a comprehensive knowledge.
ICVs can check the aggregated  knowledge that is stored at their surrounding RSUs, and exploit the comprehensive knowledge to make decisions.

(4). Cross-Region Knowledge Sharing: For an ICV driving across the regions, the RSUs will decide whether to deliver the cross-region knowledge to the vehicle for cross-regional sharing.  The impact scope $s$ of the knowledge is used for the delivery decision. The idea is that the knowledge with a small impact range does not need to be transmitted to nearby regions.  It is noted that the RSUs will transmit to cross-regional ICVs not the total ledger, but the cross-region knowledge that is encapsulated as sites.
\begin{figure}[t]
\centering
\includegraphics[scale=0.35]{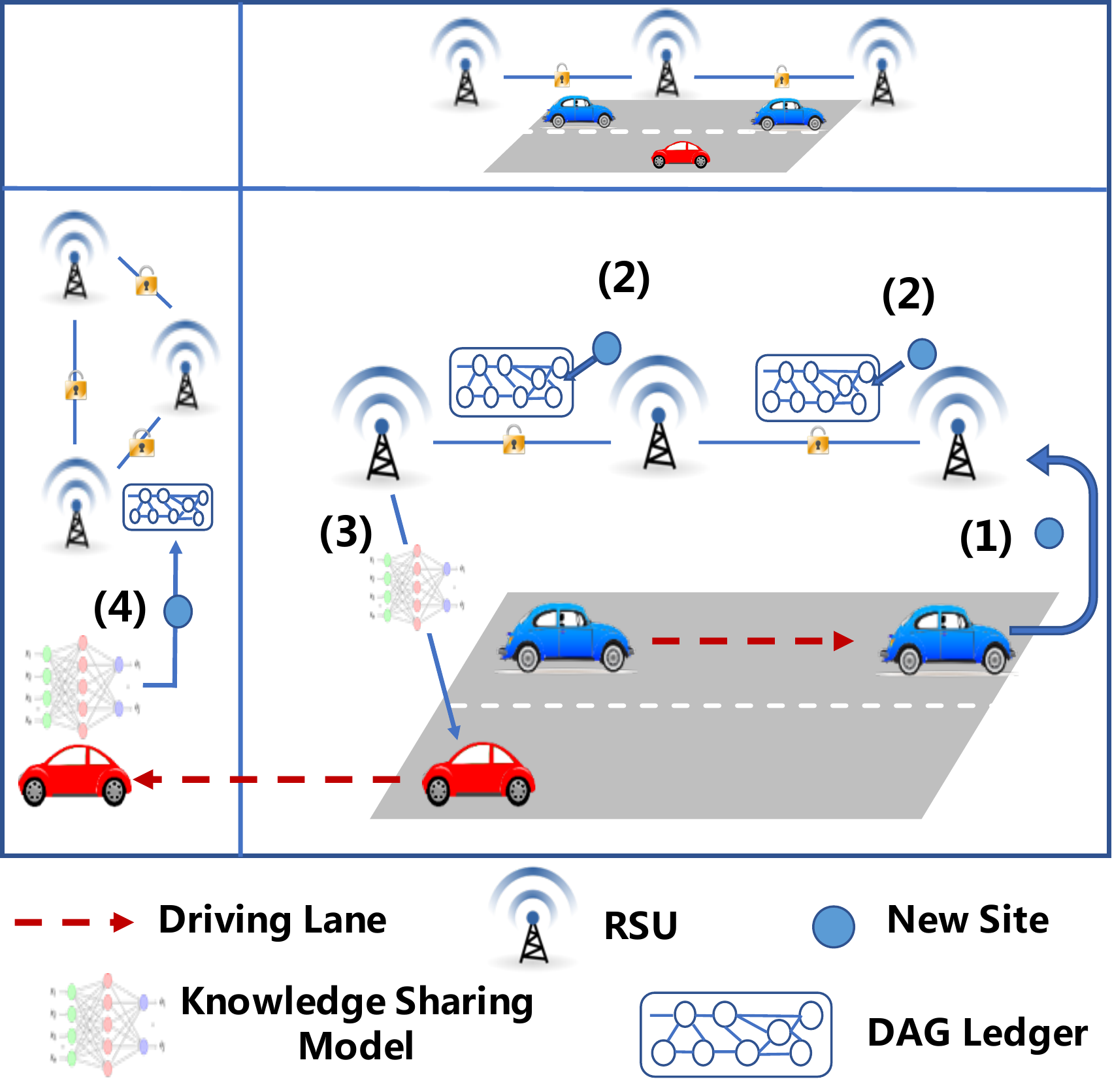}
\caption{Illustration of the Knowledge Sharing Framework}
\end{figure}

There are two characteristics of the proposed framework: (a) for local sharing, the RSUs have two identities at the same time: the DAG ledger maintainers and the ADL servers.  The RSUs can load the knowledge cached in the ledger, and implement corresponding updating schemes according to different applications, which strengthen the cooperation between the independent, distributed knowledge in DAG ledger.  
Compared with conventional ADL updating schemes, the comprehensive knowledge obtained by RSU is updated with the tamper-proof distributed knowledge that is cached in the ledger, which is essentially reliable and traceable. This can prevent the single point of failure and malicious ADL servers, since other RSUs or ICVs can check the tamper-proof ledger to verify the comprehensive knowledge.  (b) for cross-region sharing, existing blockchain based systems utilize global consensus mechanism to realize the cross-region sharing that will introduce heavy communication cost. The proposed framework enables cross-regional ICVs to carry the knowledge, and issue the knowledge as a new site in other regions. Since there is no block packing and mining process in the DAG based framework, the new site with cross-region knowledge can be directly appended of other regions' ledger, which realize fast sharing of cross-region knowledge.

\section{DAG based Knowledge Sharing For Autonomous Driving Application}

In order to show the potentials of the framework, we investigate an application of the proposed DAG based knowledge sharing for autonomous driving, in which ICVs share their driving-control related knowledge via trained models with others, and other ICVs can exploit the shared knowledge to make better self-driving decision.

The driving-control knowledge is related to control of multiple factors such as longitudinal motion, yaw dynamic and driving route.  We will base our investigation on the open source autonomous driving environment, Airsim,  which was developed by Microsoft AI \& Research team \cite{Airsim}. In the Airsim environment, driving control is characterized with three parameters: vehicular throttle $a$, wheel steering $b$ and braking $c$.  According to the Airsim project, we define a driving style indicator (denoted by $m$) to quantify the driving style, which is computed as
\begin{equation}
m=\frac{a(1-c)+b^2}{2},
\end{equation}
The driving style indicator can reflect the following facts: whether throttle $a=0$ or braking $c=1$, the ICVs will remain stationary,  and either a large $a$ or a sharply steering of $b$ will give a large value of driving style indicator $m$.

\subsection{Workflow of the DAG based Knowledge Sharing For Autonomous Driving}

For the blockchain knowledge sharing supported autonomous driving, the following workflow is designed.

\begin{enumerate}
\item Local Training of an autonomous driving model (adM): Each ICV utilizes its on-board sensing and computing resources to train adM.
In our specific scenario, a convolutional neural network is utilized as the adM, with collected context images and its driving style indicator $m$ as input, and driving control values as output.
\item Site Packaging and Issuing: As indicated in Secion III. (1), the knowledge is encapsulated as  $S=\{H, \Phi, m,  c, \textbf{w}, Sig_v\}$, where $\Phi$ is the driving-control model adM.  Then, the site is directly append on the DAG ledger that is cached by RSUs.  
\item TSA-based Site Verification: As indicated in Section III.(2), the issuing ICV chooses two existing tips on the DAG ledger according to TSA, and utilizes its test dataset to verify the adMs encapsulated in the two tips. We utilize the testing accuracy $l$ to quantify the performance of the trained adMs \cite{learning_quality}. In the autonomous driving application,  the testing accuracy can be denoted by the control gap between the real  control parameters of testing data and the predict parameters, which can be measured by  absolute error (denoted by $e$) as
\begin{equation}
e=\frac{1}{n}\sum_{i=1}^N|y_i-f(x_i)|,
\end{equation}
where $f(x_i)$ is the predict control parameter of the trained adM, and $y_i$ is the actual control of the testing data. If the testing performance gap is smaller than a certain value,  $i.e.$ $e \leq \epsilon$, the verification process is deemed as valid, and the ICV can append the new site on the DAG ledger.

\item Knowledge Sharing and Application to Autonomous Driving: After the verification process, other ICVs can check and download the shared adM that is recorded in the DAG ledgers, then they can input their own image and driving style indicator to the updated adM model to produce control output for autonomous driving.
\end{enumerate}

\subsection{Reversed Two-hop Tip Selection and Driving-style Classified Ledger}

While the DAG blockchain system has good security performance against malicious attacks, 
it has weakness in the application to autonomous driving control. 
In the existing top DAG blockchains such as Hashgraph, IOTA utilizes random-walk based TSA, which uses an approach of randomly verifying the tips for new sites.
In the autonomous driving scenarios where the driving models are related to driving style, the "random verification" manner will degrade the verification rate and the successful rate of new sites being recorded to the chain, thus degrade the knowledge sharing efficiency.

In order to solve the above problem, we design a new TSA for the autonomous driving application, which is called reversed two-hop tip selection algorithm (RTH-TSA). The RTH-TSA algorithm enables the new sites to verify the tips with similar driving style.

\textbf{RTH-TSA:} Specifically,  we reverse the tracing order of site to develop fast tip selection. 
Traditional DAG utilizes the Markov chain Monte Carlo (MCMC) algorithm to reach consensus, in which the vehicles is required to trace and verify the previous $W$ sites (parameter $W$ is called particle deep and its value is large \cite{tangle}).  In tour proposed RTH-TSA,
when a new site chooses tips to append, it does not have to trace from particle deep. Conversely, it directly chooses two existing tips from its side to connect.  Accordingly, the selection probability from site $x$ to $y$ can be shown as:
\begin{equation}
P_{xy}=\frac{exp\{-\alpha(CW^1_y+CW^2_y)-\beta (m_x-m_y)^2\}}{\sum_{z \in \mathbb{T}}  exp\{-\alpha(CW^1_z+CW^2_z)-\beta (m_x-m_z)^2\}},
\end{equation}
where $\mathbb{T}$ represents the set of current tips of the DAG ledger and $\alpha$, $\beta$ are both positive weight parameters. $CW^1_y$ and $CW^2_y$ are cumulative weights of two sites that are approved by tip $y$.  The cumulative weight adopted in Eq. (3) aims to prevent $lazy$ sites from choosing two old sites instead of two tips. If the lazy sites are appended to the existing old sites, the probability of the lazy sites being appended will be largely reduced and they will eventually become the "orphan" sites.

\begin{figure}[t]
	\centering
	\includegraphics[scale=0.31]{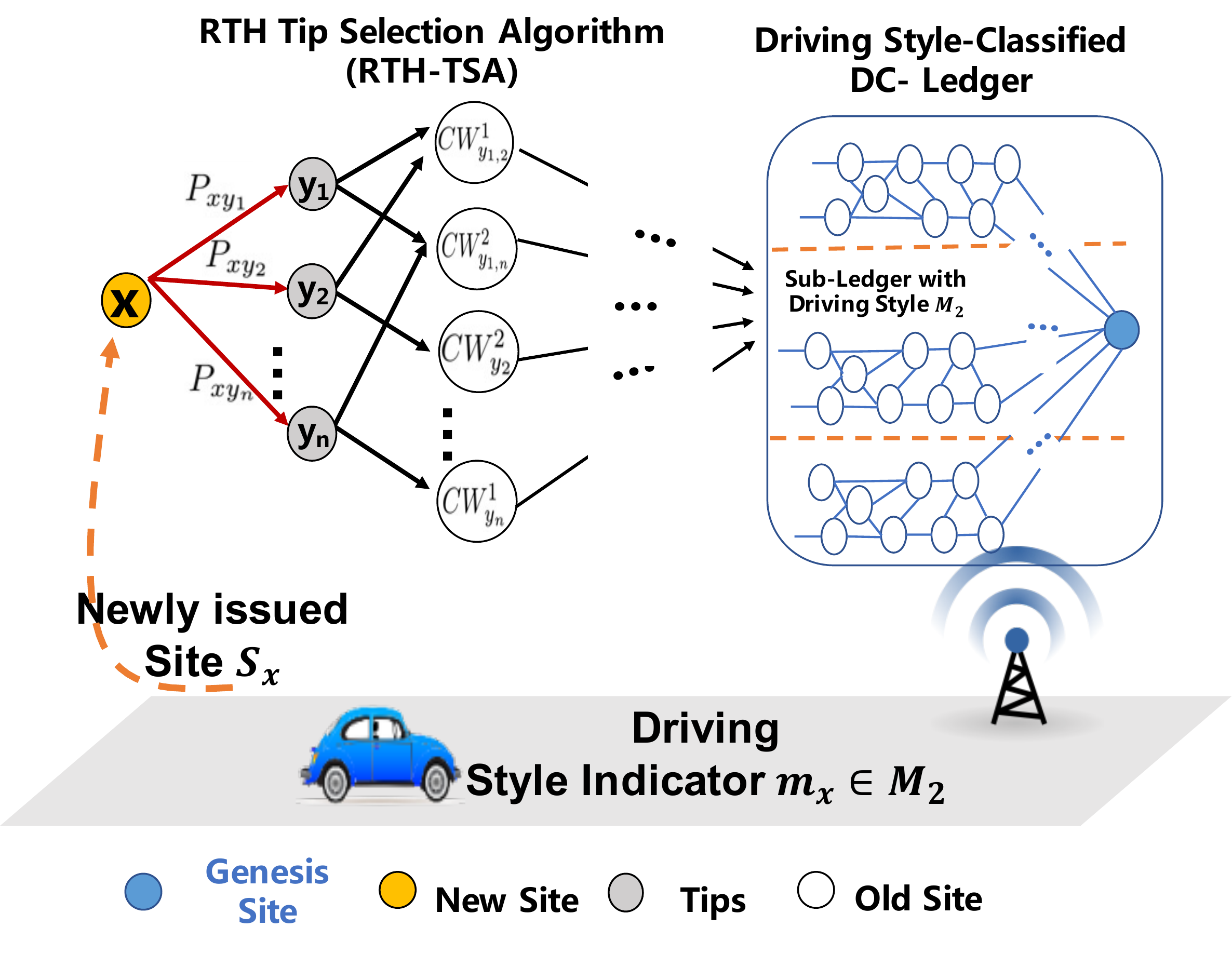} 
	\caption{Illustration of RTH-TSA Algorithm and DC-Ledger}
\end{figure}

\textbf{DC-Ledger:} Since the tip selection algorithm defines how a new site connects to the DAG ledger, $i.e.$, how the site is verified by others. 
The TSA will determine the ledger structure and how the sites are appended on the DAG. In this case, the proposed RTH-TSA enforces the new sites to verify and append behind those tips with similar driving styles. The total DAG ledger will eventually evolve to multiple sub-ledger,each of which records the sites with similar driving styles.  This realizes the driving-pattern classification of the ledger (DC-ledger). According to the verification process described in subsection-A, the verification rate can be enhanced. The design of the RTH-TSA and DC-Ledger is illustrated in Fig. 2.

\subsection{Knowledge Sharing among Multiple Regions}

In the proposed framework, we utilize moving ICVs to exchange the knowledge among different traffic regions, rather than using global consensus. 
Note that there is no need to transmit all the knowledge to other regions. 
The knowledge with a small impact range will not be transmitted to other regions, as the knowledge does not affect other regions. 
The RSUs only have to transmit those important adM knowledge to the adjacent regions to the cross-regional ICVs.

To ensure trust and reliability of the shared knowledge for the  autonomous driving application, 
cross-regional knowledge needs to be carried by authenticated ICVs. 
Therefore, it is crucial to design an identity verification scheme for those cross-regional ICVs.  
Existing schemes utilize identity tokens for the verification process \cite{token1},  which will introduce extra storage consumption and communication cost for knowledge sharing. 
Therefore, we design an efficient identity verification scheme for cross-regional ICVs that does not use identity tokens.

\emph{\textbf{Fast Identity Authentication Scheme:}}  There are two special sites during the authentication process: 
one is cross-regional site $S^{c}_r$, and the other is \emph{IdentityStone} $S_{is}$.  
As illustrated in Fig. 3, considering a cross-regional scenario that ICV $y$ leaves region $A$ to region $B$.
If RSUs of region $A$ decide to share the knowledge with $B$, 
they send to $y$ the cross-regional site $S^{c}_r$, containing 
the knowledge and the current scope $s$.
Meanwhile,  RSUs of region $B$ will issue a \emph{IdentityStone} $S_{is}$, which is signed with its private key $Sig_r$. The formats of the two special sites are expressed as
\begin{equation}
\begin{aligned}
&S^{c}_r=\{H, K^s, \textbf{w}=0, Sig_r\}\\
&S_{is}=\{H, \textbf{w}=0, Sig_r\} 
\end{aligned}
\end{equation}

\begin{figure}[t]
	\centering
	\includegraphics[scale=0.32]{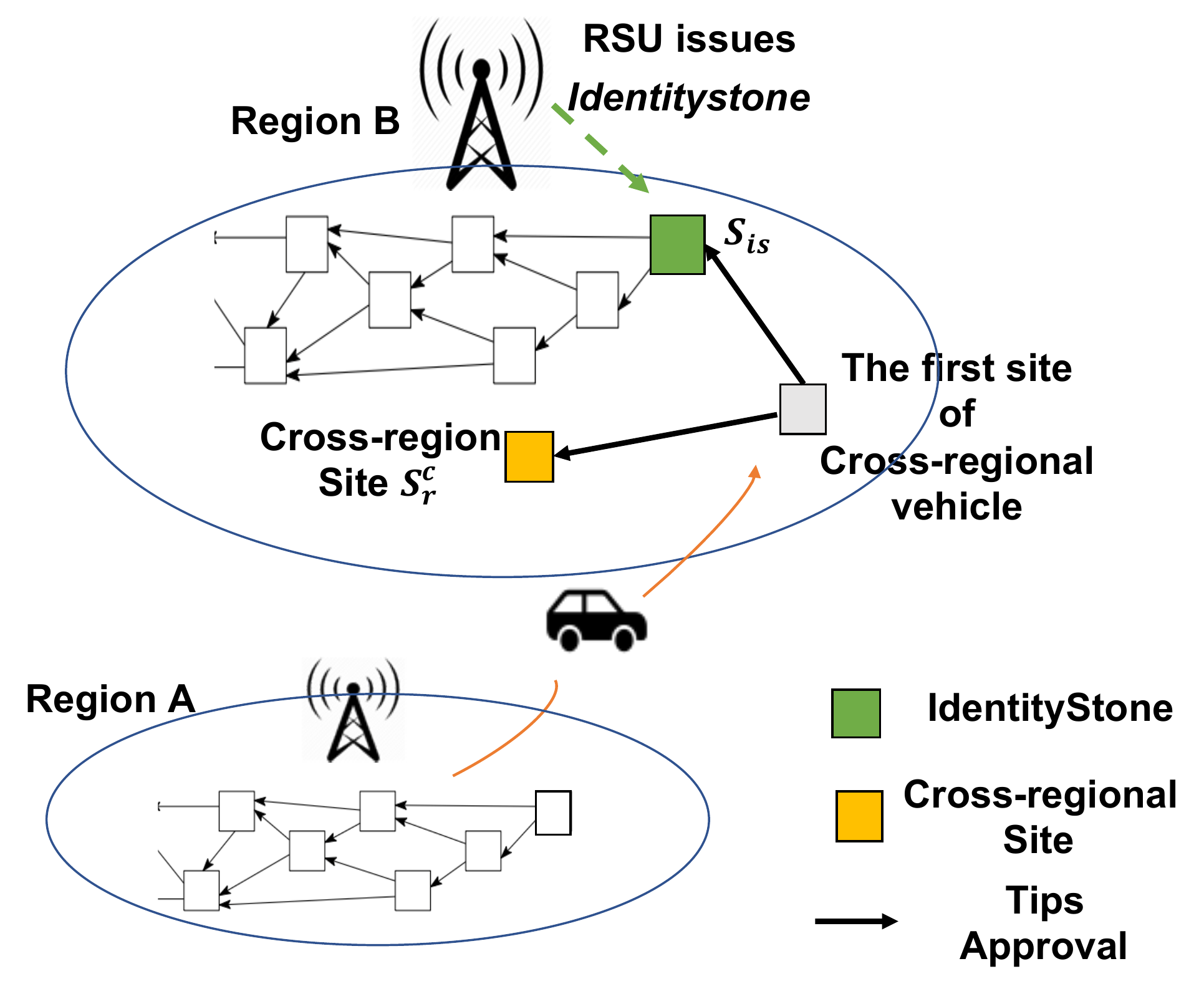} 
	\caption{Fast Authentication for Cross-regional Vehicles}
\end{figure}

To make the scheme lightweight, we remove some unnecessary elements in IdentityStone such as knowledge $K$ and address. 
The reason for setting the cumulative weight and own weight to 0 is to avoid affecting the current weight of the DAG ledger,
which is inspired by the Project $Tangle$ \cite{tangle}.

When a vehicle $y$ issues the first site in the new region $B$, it must approve the two specific sites $S_{is}$ and $S^c_r$. 
After that,  $y$ is deemed to have completed identity verification. 
Noting that the cumulative weight of both sites is 0, the newly coming sites of other ICVs in region $B$ will first verify the site of the cross-regional vehicle $y$ according to Eq.(3), 
which completes the fast authentication of vehicle $y$.

As $y$ will approve $S^c_r$ from its original region $A$ when it joins a new region, vehicle $y$ is traceable if it is malicious.  
The scheme utilizes the process of site approval that is inherent in DAG and does not need extra storage space, which realizes lightweight process.


\subsection{The Ledger Convergence}
To tackle the problems in the existing TSA of DAG, we exploited the architecture of DC ledger with multiple sub-ledgers. 
However, to ensure the feasibility and effectiveness of the DAG system, the ledger must remain convergent, 
which means that the total number of tips of the DC ledger cannot escape to infinity.
In this subsection, the convergence of the ledgers will be analyzed.


Assume that the ICVs issue new sites independently with the distribution $I_n$. Let us denote $T_n$ as the number of tips in the proposed DAG at the $n$-th approval round. In this case, the total number of tips in the network can be expressed as

\begin{equation}
T_n=T_{n-1}+I_{n}-A_n,
\end{equation}
where $A_n$ is the number of approved tips by $I_n$. Assume that the driving style indicator range is divided into $K$ discrete intervals $\mathbb{M}=\{M_1,...,M_k,...M_K\}$, and the mobility distribution of vehicles is $m(t)$ with respect to time $t$. For one specific sub-ledger $k$ that locates on the mobility interval $M_k \in \mathbb{M}$, the expectation of $T_n^{M_k}$ is
\begin{equation}
E[T_n^{M_k}]=E[T_{n-1}^{M_k}]+E[I_n^{M_k}]-E[A_n^{M_k}].
\end{equation}

In order to obtain the value of $E[T_n^{M_k}]$, the distribution of $I_n^{M_k}$ and $A_n^{M_k}$ should be specified.  The incoming tips can follow the  poisson, uniform and gamma distributions, $etc$. In this section, we only select one ordinary poisson distribution with a incoming rate $\lambda$ and prove the convergence under the distribution. More distribution examples will be investigated in the simulation part. Hence, the Eq. (7) can be expressed as $E[T_n^{M_k}]=E[T_{n-1}^{M_k}]+\lambda h P\{m(t) \in M_k\}-E[A_n^{M_k}]$, where $h$ is the average time of each approval. .

To obtain the value of $E[A_n^{M_k}]$, the joint probability $P\{ A_n^{M_k}(h)=n_k, k=1,2...,K\}$ during time interval $(0, h)$ should be computed. Note that for $n_k$ tips with driving style indicator $M_k$, there must be a total of $\sum_{k=1}^{K} n_k$ tips during $(0, h)$. Therefore, conditioning on total tips yields
\begin{equation}
\begin{aligned}
&P\{A_n^{M_1}(h)=n_1, A_n^{M_2}(h)=n_2..., A_n^{M_K}(h)=n_K\} \\
&=P\{A_n^{M_1}(h)=n_1..., A_n^{M_K}(h)=n_K|N(h)=\sum_{k=1}^{K} n_k\} \\
&\times P\{N(h)=\sum_{k=1}^{K} n_k\}
\end{aligned}.
\end{equation}

For an arbitrary incoming site $x$, the probability $P_{xy}^{M_k}(s)$ that site $x$) approves the tip $y$ with mobility $k$ at time $s$ is shown in Eq. (3). 
For simplicity, the cumulative weight $CW$ in (2) can be ignored because various sites have similar weights. 
Therefore, $P^{M_k}(s) \approx \frac{exp\{-\beta[M_k-m(s)]^2\}}{\sum_{l: M_l \in \mathbb{M}} exp\{-\beta [M_l-m(s)]^2\}}$. 
According to Theorem 5.2 in \cite{Probability}, tip approval occurs at a certain time uniformly distributed in $(0,h)$. 
Therefore, the probability that one site will approve those tips with driving style indicator $M_k$ is computed by
\begin{equation}
P^{M_k}=\frac{1}{h} \int_{0}^{h} P^{M_k}(s) ds
\end{equation}
and is independent of the other tip approval events. 
Hence, the conditional probability is a multinomial distribution with parameters $P^{M_1}, P^{M_2}..., P^{M_K}$, 
and the joint probability can be expressed as
\begin{equation}
\begin{aligned}
&P\{A_n^{M_1}(h)=n_1, A_n^{M_2}(h)=n_2..., A_n^{M_K}(h)=n_K\} \\
&=\frac{(\sum_{k=1}^{K} n_k)!}{n_1!n_2!...n_K!}(P^{M_1})^{n_1}(P^{M_2})^{n_2}...(P^{M_K})^{n_K} e^{-\lambda h}\frac{(\lambda h)^{\sum n_k}}{(\sum n_k)!} \\
&=\prod_{k=1}^{K} e^{-\lambda h P^{M_k}} \frac{(\lambda h P^{M_k})^{n_k}}{n_k!}.
\end{aligned}
\end{equation}

From the above analysis, it can be obtained that the variable $A_n^{M_k}(h)$ follows a Poisson distribution with parameter $\lambda h P^{M_k}$. Therefore, the expectation $E[A_n^{M_k}(h)]$ is computed by:
\begin{equation}
E[A_n^{M_k}(h)]=\lambda \int_{0}^{h} \frac{exp\{-\beta[M_k-m(t)]^2\}}{\sum_{l: M_l \in \mathbb{M}} exp\{-\beta [M_l-m(t)]^2\}}dt.
\end{equation}

By substituting $E[A_n^{M_k}(h)]$ into Eq. (11), the expectation of $T_n$ can be obtained
\begin{equation}
\begin{aligned}
&E[T_n]=\sum_{k=1}^K E[T_{n}^{M_k}]=E[T_{n-1}]+\sum_{k=1}^K (E[I_n^{M_k}]-E[A_n^{M_k}]) \\
&=E[T_{n-1}]+\lambda h \cdot 1-\lambda \int_{0}^{h} 1 dt =E[T_{n-1}].
\end{aligned}
\end{equation}

Consequently, during each approval round, the total number of tips of all sub-ledgers is the same as that of the previous round. Although the number of tips of each sub-ledger in each round may be different, the total number of their tips remains constant, which proves the convergence of the ledger.

\section{Asynchronous Distributed Learning enhanced DAG For Autonomous Driving}

Although the proposed DAG based knowledge sharing can provide reliability and security for autonomous driving application, there could still be accuracy and communication related issues: 
(a). As the adMs cached in the DAG ledger are only the local models that are trained independently, they could be far away from the optimal one obtained by joint training; 
(b). For the large-scale autonomous driving scenarios, the ICVs continuously broadcasting their adMs could significantly increase the communication load of the ICV networks.  

To tackle the above issues, we exploit asynchronous distributed learning (ADL) to enhance the DAG based knowledge sharing system,
in which the RSUs will be responsible for the processing as well as maintaining the adMs in the DAG ledger.

\subsection{Adaptive Asynchronous Distributed Learning Enhanced DAG Knowledge Sharing}
The design objectives of the ADL algorithm is to exploit the correlation among different adMs in the DAG ledger, 
and obtain a comprehensive global model to enhance the accuracy of the autonomous driving related model. 
Compared with conventional synchronous distributed learning, the ADL can support ICVs to upload their local models in an asynchronous way,  
and they can obtain the on-the-fly updated model without waiting for other ICVs, which is suitable for the high dynamic autonomous driving scenarios.
Furthermore, considering the difference of the ICV training ability, there is no need to enforce all the ICVs to upload their local adMs.  
Some local adMs with lower learning performance may even degrade the accuracy of global model. 
Therefore, we propose an adaptive scheme for the ADL, aiming to enhancing the accuracy of final global model and reduce the communication cost among ICVs. 

The specific workflow of the adaptive ADL enhanced DAG knowledge sharing process is described as follows.

\begin{enumerate}
\item Initialization: The RSUs cache a global adM (GadM) in advance
with model version $V_G=1$, and the initial GadM can be obtained by  offline learning methods. Each RSU caches an identical and periodically updated test dataset.

\item GadM Broadcasting: The RSUs periodically broadcast the reference gap $e_G$ of current GadM that is evaluated by RSUs' test dataset as indicated in Section IV-A-(3), current model version $V_G$, together with its digital signature $Sig_r$ to surrounding ICVs.

\item Judging the Reliability of GadM: Upon receiving the multiple information from RSUs, ICVs check the received versions and signatures of GadM, which can prevent the single point failure of RSUs. The ICVs will choose the version sent by most RSUs as the credible GadM version.

\item Adaptive adM Site Issuing: After verify the reliability of the GadM,  the ICV will compare the test gap $e_v$ of its own local adM with reference gap $e_G$. If $e_v \leq e_G$,  the ICV can believe that its local adM outperform the current GadM, and it will encapsulate its model and its driving styles indicator $m$ into new site $S$, as indicated in Section IV-A. If $e_v> e_G$, the ICV will not issue a new site to the DAG, and will download the current GadM from RSUs as its model.
  
\item Asynchronous Updating GadM: By receiving the new adM from the ICV, the RSUs will first utilize its test dataset to evaluate the received adM. If the test gap $e_r \leq \epsilon$ ($e_r$ is the test gap of the received adM tested by RSUs), the adM is deemed as valid and can be appended by other new sites.  Then,  the RSUs update the current version $V=V+1$, and update the GadM in an asynchronous way by referring to the driving style based updating scheme, which will be discussed in the next subsection. If $e_r > \epsilon$,  the RSUs will send a fail signaling back to the ICV, and back to step (2).
\end{enumerate}

Noting that the adaptive site issuing process utilizes the same metric, which is the test performance gap $e$ defined in Eq.(3), to measure the model quality.
The metric is used to facilitate the system implementation and management in practice. 
With the adaptive scheme,  those adMs with a low accuracy will not be appended on the DAG ledger, 
which will enhance the quality of global model GadM. Furthermore, with less uploaded models  the communication loads between the ICVs and the RSUs can be significantly reduced.

\subsection{Driving Style based Asynchronous Updating Scheme}


Next, we will elaborate the process of \emph{Asynchronous Updating GadM} in Step 5) of the adaptive ADL presented in the last subsection.  
Taking into account the variety in the driving styles of ICVs,  there are also differences in accuracy between adMs of ICVs. It is inappropriate to aggregate the adMs regardless of their driving styles.  Therefore, we propose a driving style based updating scheme for the ADL algorithm, which combines both driving styles of ICVs and the freshness of adMs.

Considering the ADL process that $N$ ICVs succeed in issuing adMs within $T$ time, each ICV utilizes stochastic gradient descent (SGD) algorithm to implement local training  $\theta^{l}(t)=\theta^{l}(t-1)-\gamma \triangledown F^{l}(t)$, where $\theta^l$ represents the adM and $F^l$ is the local optimization function. Then, the global model GadM $phi$ at time $t$ can be expressed as 
\begin{equation}
\Phi(t)=(1-\alpha_t f_t)\Phi(t^-)+\alpha_t f_t \theta^l_t
\end{equation}
where $t^-$ is defines as the time index of last asynchronous updating, and $f_t$ is the function of computing the freshness of local models 
that is utilized to describe the decay of old weights over time. 
The freshness (denoted by $f_t$) is computed as
\begin{equation}
f_t=e^{1-\frac{t}{T}},
\end{equation}
and the $\alpha_t$ in Eq.(14) represents the driving-related weight.

Driving style is a reflection of the interaction between ICVs and the external environment, which contains the knowledge of context and driving control. 
For example, on the same road segment, ICVs with different driving velocities will have diverse safe braking distance. Therefore, it is inappropriate to aggregate all adMs together in one global model regardless of the impact of driving styles.  In this case, we define $\alpha_t$ as
\begin{equation}
\alpha_t= 1-\widetilde{m_t}
\end{equation}
where term $\widetilde{m_t}$ is the relative driving style indicator of ICV at time $t$,  that is described as $\widetilde{m_t}=|m_t-\overline{M_t}|$, and $\overline{M_t}$ is the average driving style indicator, $m_t$ is the driving style indicator defined in Eq.(2).  

The driving style based updating process is illustrated by \textbf{Algorithm 1}. The computational complexity of the proposed algorithm is mainly determined by the process from Line 2 to Line 13. According to \cite{algorithmO}, the time complexity of sampling and updating from Line 7 to Line 9 is $O(logN_{tree})$, where $N_{tree}$ is the number of nodes in the sum-tree used to compute the priority probability. 
As the time complexity of model aggregation scheme in Line 11 to Line 13 is $O(1)$, the total  time complexity is $O(N logN_{tree})$.

It can be found that the proposed updating scheme differs from the traditional DL process. The global model is updated with an adjustable weight $\alpha_t f_t$, which is related to the freshness of adMs and driving style indicator of ICVs.  Next, we will analyze the optimal weight and driving style of ICVs under the proposed updating scheme. Simulations also demonstrated the proposed scheme outperformed existing DL algorithms in the autonomous driving scenarios.

\begin{algorithm}[t]
  \caption{Driving Style Based Asynchronous Updating Scheme}  
  $\textbf{Initialization:}$ $N$ ICVs with driving style indicators $\{m_1,..., m_N\}$,   Global Training Time $T$ \\
  \textbf{for} $n = 1: N$\\
    \ \ \ \textbf{Local adM Training:} \\
    \ \ \ $\theta^l_n \gets \Phi$\\
    \ \ \ $B_n$ $\gets$ Local training batch of vehicle $n$  \\
  \ \ \ \textbf{for} each local epoch in $L$ \\
  \ \ \ \ \ \ \ \textbf{for} batch b $\in$ $B_n$ \\
  \ \ \ \ \ \ \ \ \ \ \ training model $\theta^l_n$ $\gets$ $\theta^l_n - \gamma \bigtriangledown F^l(\theta^l_n|b)$\\
 \ \ \ \textbf{return} Local model $\theta^l_n$, Local Completion time $t_n$\\
  \ \ \ \textbf{Model Aggregation:} \\
  \ \ \ $f_{t_n} \gets e^{1-\frac{t_n}{T}}$\\
  \ \ \  $\alpha_{t_n}=e^{-\widetilde{m_{t_n}}}$\\
  \ \ \ $\Phi(t)=(1-\alpha_{t_n} f_{t_n})\Phi(t^-)+\alpha_{t_n} f_{t_n} \theta^l_t$\\
 \textbf{Output} Final global model $\Phi(T)$
\end{algorithm}

\subsection{Analysis of Optimal Model Weight and ICV Driving Style}

Since we adopt a driving style based updating scheme, it is essential to discuss the impact of the weight on the training accuracy.  
The convergence bound of distributed learning is utilized, which characterizes the gap between the optimal performance and the training performance. 
The convergence bound analysis of existing ADL algorithms has been widely discussed \cite{convergence1},  
and the specific proof of convergence is similar with the pioneer work \cite{ADLcon}. 
For the sake of limited space, we will not present the proof but directly give the convergence bound of the proposed ADL.

\emph{Convergence Bound}: 
Under two assumptions that for all \textbf{v} and \textbf{w}:
\emph{L-smooth:}  f(\textbf{v})-f(\textbf{v}) $\leq$ (\textbf{v} - \textbf{w})$^{T} \bigtriangledown$ f(\textbf{w})+$\frac{L}{2}$  ||\textbf{v} - \textbf{w}||$_2^2$,
\emph{$\mu$-strong convex:}:  f(\textbf{v})-f(\textbf{v}) $\geq$ (\textbf{v} - \textbf{w})$^{T} \bigtriangledown$ f(\textbf{w})+$\frac{\mu}{2}$  ||\textbf{v} - \textbf{w}||$_2^2$,
the proposed ADL within $T$ global epoch converges to a critical point:
\begin{equation}
\begin{aligned}
&\mathcal{P}=\min \limits_{t=0}^{T-1} \mathbb{E}||\bigtriangledown F(\theta_t)||^2 \leq \\
&  \mathcal{O}\left(\frac{1}{\overline{\alpha} \gamma \epsilon TH_{min}}+\frac{\gamma H_{max}^3 +\overline{\alpha} KH_{max}+\overline{\alpha}^2 \gamma K^2H_{max}^2 +\gamma K^2 H_{max}^2 }{\epsilon H_{min}}\right)
\end{aligned}
\end{equation}
where $\overline{\alpha}$ is the expectation of driving style indicator based weight  within time duration $(0,T)$, $i.e.$ $\overline{\alpha}=\frac{1}{N_T}\sum_{t=1}^{N_T} \alpha_t$, where $N_T$ is the total number of ICVs in $T$ time. 
$\gamma$ is the learning rate, $\epsilon >0$ is a small constant. $H_{min}$ and $H_{min}$ are the local minimum and maximum iteration rounds, respectively. 
$K$ is the maximal interval between two updating of ICV models, $i.e.$ $t-t^- \leq K$.

Now we can utilize $\mathcal{P}$ to determine the optimal weight $\overline{\alpha}^*$.  
We approximately simplify the convergence bound as $\mathcal{P} \approx \frac{1}{\overline{\alpha} \gamma \epsilon TH_{min}}+\frac{\gamma H_{max}^3 +\overline{\alpha} KH_{max}+\overline{\alpha}^2 \gamma K^2H_{max}^2 +\gamma K^2 H_{max}^2 }{\epsilon H_{min}} $. 
Therefore, the bound can be expressed as a function of weight $\overline{\alpha}$.
Letting $x$ denote $\overline{\alpha}$, the bound can be transformed as 

\begin{equation}
\mathcal{P}(x)=\frac{1}{Ax}+Bx+Cx^2+D
\end{equation}
where $A=\frac{1}{\gamma \epsilon TH_{min}}$, $B=\frac{K \delta}{\epsilon}$, $C=\frac{\gamma K^2 \delta H_{max}}{\epsilon}$, $D=\frac{\gamma \delta H_{max}^2+\gamma \delta K^2H_{max}}{\epsilon}$, $\delta=\frac{H_{max}}{H_{min}}$.

By obtaining the first derivation of $\mathcal{P}(x)$, it yields
\begin{equation}
\mathcal{P}'(x)=-\frac{1}{Ax^2}+2CX+B
\end{equation}

Since $A >0$, by letting $\mathcal{P}'(x)=0$, we can obtain a cubic equation
\begin{equation}
2ACx^3+ABx^2-1=0
\end{equation}

Hence, the extreme values of function $\mathcal{P}$ can be obtained by solving Eq. (20).
In order to solve the cubic equation and obtain a feasible solution, Cardano's formula  \cite{Cardano} is utilized and Proposition  1 is proposed.

\emph{\textbf{Proposition 1}: The convergence bound $\mathcal{P}$ reaches its minimal value at point}
\begin{equation}
\overline{\alpha}^*=\sqrt[3]{\frac{\epsilon^2 T}{2\delta^2K^2}}
\end{equation}
if $\gamma^3=\frac{1}{27T K\epsilon^2 \delta H^3_{min}}$.

\emph{Proof:}
For an arbitrary cubic equation $aX^3+bX^2+cX=d=0$, we define three auxiliary variables
\begin{equation}
\begin{aligned}
&A_0=b^3-3ac \\
&B_0=bc-9ad \\
&C_0=c^2-3bd
\end{aligned}
\end{equation}

By substituting the equation above into Eq.(20), a discriminants can be obtained that 
\begin{equation}
\begin{aligned}
&\bigtriangleup=B_0^2-4A_0C_0=12A^2(27C^2-AB^3) \\
&=12A^2(\frac{27\gamma^2 T^4 \delta^2 H_{max}^2}{\epsilon^2}-\frac{T^2\delta^3}{\gamma \epsilon^4 H_{min}})
\end{aligned}
\end{equation}

According to \cite{Cardano}, there are one real root and two equal real roots if $\bigtriangleup=0$. 
By letting $\bigtriangleup=0$, it can be obtained that 
\begin{equation}
\gamma^3=\frac{1}{27\epsilon^2 \delta H_{min}^3TK}
\end{equation}
And an auxiliary parameter is defined as 
\begin{equation}
\begin{aligned}
&\mathcal{K}=\frac{B_0}{A_0}=\frac{18AC}{A^2B^2}\\
&=\frac{\frac{18\gamma T^2 \delta H_{max}}{\epsilon}}{\frac{T\delta^2}{\gamma \epsilon^3H_{min}}}=18\gamma^2\epsilon^2H^2_{min} T
\end{aligned}
\end{equation}
Then, the three real roots of Eq.(20) can be represented as 
\begin{equation}
\begin{aligned}
& X_1=\mathcal{K}-\frac{AB}{AC}=-\frac{\frac{K \delta}{\epsilon}}{\frac{2\gamma K^2 \delta H_{max}}{\epsilon}}=\mathcal{K}-\frac{1}{2 \gamma T H_{max}}\\
&\ \ \ =\mathcal{K}-\frac{27}{2}\epsilon^2 \gamma^2H^2_{min}T=18\gamma^2\epsilon^2H^2_{min}T-\frac{27}{2}\epsilon^2 \gamma^2H^2_{min}\\
&\ \ \ =\frac{9}{2}\gamma^2\epsilon^2H^2_{min}T= \sqrt[3]{\frac{\epsilon^2 T}{2\delta^2K^2}}\\
&X_2=X_3=\frac{-\mathcal{K}}{2}=-9\gamma^2\epsilon^2H^2_{min}T\\
\end{aligned}
\end{equation}

Obviously $X_2=X_3 < 0$, since $\overline{\alpha}$ is the expectation value of vehicular mobility, and $\overline{\alpha} > 0$, thus $\overline{\alpha}=X_1$. 
By substituting $X_1$ into the second derivation of $\mathcal{P}(x)$, and $\mathcal{P}''(X_1) > 0$.  
Consequently, the convergence bound $\mathcal{P}$ reach its minimal point at $X_1$. $\hfill \blacksquare$

Based on Proposition 1,  we have two remarks with respect to model weight $\overline{\alpha}$ and driving style indicator $m$.

\emph{Remark 1: Compare with precious GadM,  the new uploaded adM should  have a higher weight.} 
Proposition 1 gives the optimal weight $\overline{\alpha}^*$ that is associate with learning time $T$. 
With the increase of $T$, the value of $\overline{\alpha}^*$ will also increase, hence, considering a long-term training process,  
the weight of new updated model $\overline{\alpha}$ is greater than the previous model $(1-\overline{\alpha})$.

\emph{Remark 2: On the basis of reducing communication cost as well as not affecting the accuracy of final model, 
those ICVs with indicator $m$ far away from the average $\overline{M}$ should choose not to upload their local adMs.} 
As indicated in \emph{Remark 1}, the optimal $\overline{\alpha}^*$ will increase with learning time $T$. 
Since $\overline{\alpha}$ can be expressed as $\overline{\alpha}=\frac{1}{N_T}\sum_{t=1}^{N_T} \alpha_t=\frac{1}{N_T}\sum_{t=1}^{N_T}(1-|m_t-\overline{M_t}|)$, 
we can decrease the number of those ICVs with $m_t$ that are far away from the average $\overline{M_t}$, 
so that the value of $\overline{\alpha}$ will not decrease. The remark can be explained by the fact that the local models trained by those ICVs with a deviated driving style cannot reflect the accurate global context,  and we can reduce the use of models from those ICVs to reduce the uploading communications while maintaining the global training quality.

\section{Simulation Results}
In this section, we evaluate the proposed DAG based knowledge sharing framework and ADL algorithm by simulation experiments. 
The performance of the DAG based framework is first investigated, followed by simulation tests of the proposed ADL scheme.

\subsection{Simulation Setup}

We evaluate the  DAG based knowledge sharing framework in Matlab, in which the sites contains both driving style indicator $m$ and shared knowledge ($i.e.$ the adM). 
New sites are generated following three distributions: uniform distribution within (400, 500), poisson distribution with $\lambda=700$ and gamma distribution with $\alpha=200, \beta=1$. 
The tip selection probability is based on Eq. (4).
We further control ICVs by Logitech G29 Driving Force Steering Wheel and Pedal to simulate real driving style of ICVs.  
Three types of driving styles are adopted: 
$m_1$-type for ICVs has a large throttle $a=0.9$, a small steering $b \in [-0.2, 0.2]$, braking $c=0$, 
and the corresponding driving style indicator $m_1 \in [0.45, 0.47]$; 
$m_2$-type has $a=0.5$, large steering $b \in [0.6, 0.9], [-0.6, -0.9]$, $c=0$, and $m_2 \in[0.43, 0.65]$; 
$m_3$-type has $a=0.9$, $b=0.1$, high braking pattern $c \in [0.5, 0.8]$, and $m_3 \in [0.095, 0.23]$.
The tip verification process is simulated with setting $m_2$-type ICVs to train a target adM, 
 and enforcing the three types of ICVs to respectively test the target adM in terms of test gap, as described in Section IV.A.(3).

\begin{table}[t]
\newcommand{\tabincell}[2]{\begin{tabular}{@{}#1@{}}#2\end{tabular}}
\centering
\caption{SIMULATION PARAMETERS}\label{tab:tab2}
\scalebox{0.7}{
\begin{tabular}{lc}
\hline
Parameters & Value \\\hline
Number of Vehicles & 35 \\\hline
Dataset size of distributed learning & 3000  \\\hline
Driving Style Indicator $m$ & $m_1 \in [0.45, 0.47]$, $m_2 \in [0.43, 0.65]$, \\ \ &$m_3 \in [0.095, 0.23]$\\\hline
Driving Throttle, Braking $a, c$ & [0, 1]\\\hline
Driving Steering $b$ & [-1, 1]\\\hline
adM \& GadM network models & CNN \\\hline
Own weight of each site $\omega$ & 1  \\\hline
\ & Uniform within (400, 500) \\
Sites Generation Distribution & Poisson distribution with $\lambda$=700 \\ \ & Gamma distribution with $\alpha=200, \beta=1$ \\\hline
\end{tabular}}
\end{table}

\begin{figure}[t]
\centering
\includegraphics[scale=0.25]{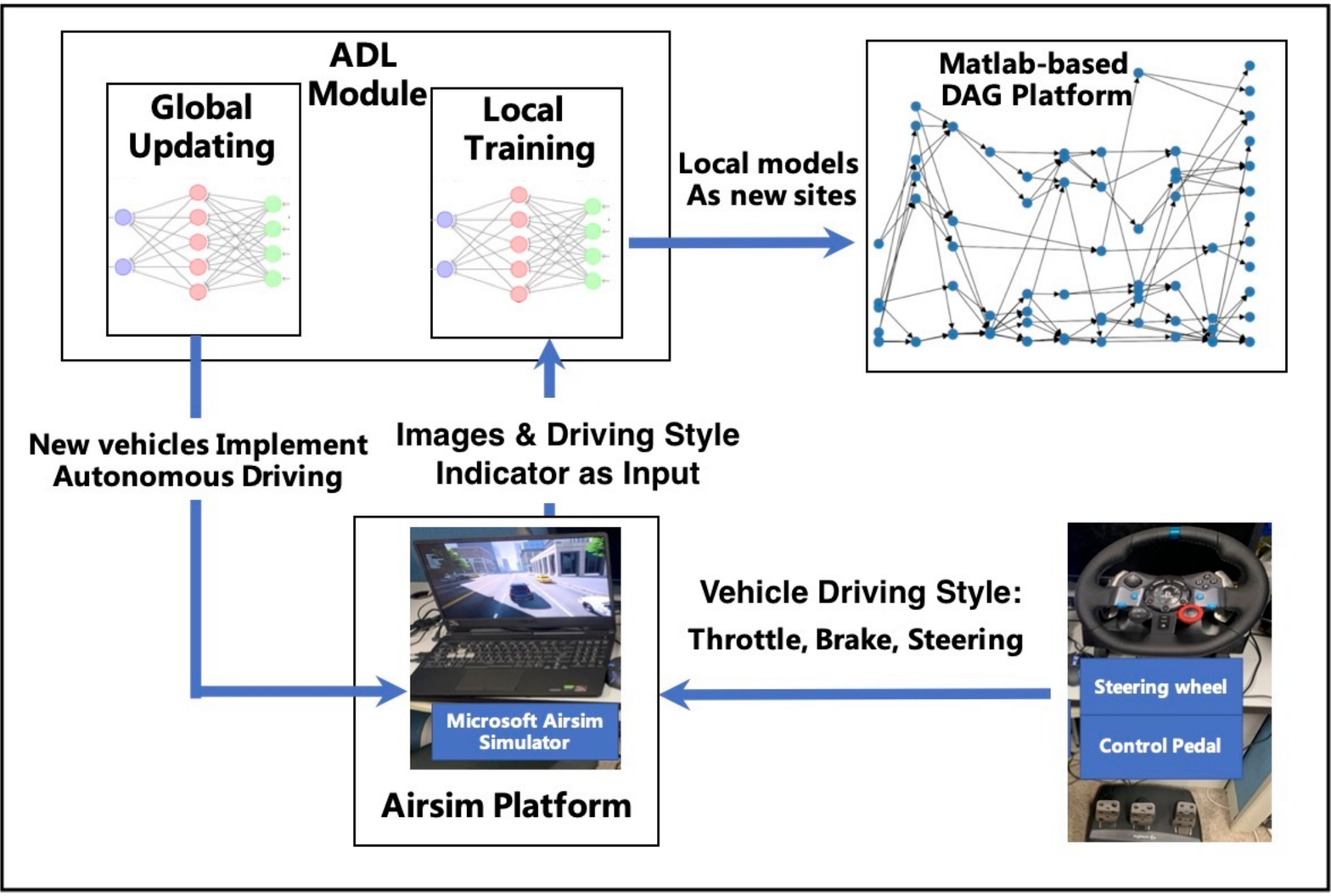}
\caption{Testing Environment}
\end{figure} 
 
Then the ADL is investigated with the open-source project Airsim, 
which provides a platform for AI research for autonomous vehicles.  
We focus on the city scenario of Airsim, in which 35 vehicles are utilized for evaluation of ADL. 
The vehicles are controlled by Logitech G29 to drive around the city and collect their surrounding images to build a dataset.  
The convolutional neural network (CNN) model is adopted as our local model.  
The driving style indicator $m$ and the collected images are chosen as the input of the CNN model, and the output is the steering value of vehicles.
The performance of ADL is evaluated in terms of loss function, test gap and violation rate.  
Here the violation rate is defined as the the percentage of bad behaviours  of new vehicles autonomous driving with the trained CNN models in Airsim, 
such as retrograde motion or driving to pedestrian lanes. 
Three existing algorithms are chosen as the comparison groups: the standard federated learning (FedAve) \cite{FedAve2} that aggregates the local models from all vehicles synchronously and ignores the driving style,  
an autonomous driving algorithm (cook-AD) proposed by Project Road Runner at Microsoft Garage \cite{cook} that only exploit the local view of vehicles without referring to other vehicles, 
and the centralized CNN that trains on the whole centralized dataset.

The simulation environment is illustrated in Fig.  4 and the main parameters used in the simulations are summarized in Table I.

\subsection{Numerical Results and Discussion}

\begin{figure}[t]
\centering
\includegraphics[scale=0.3]{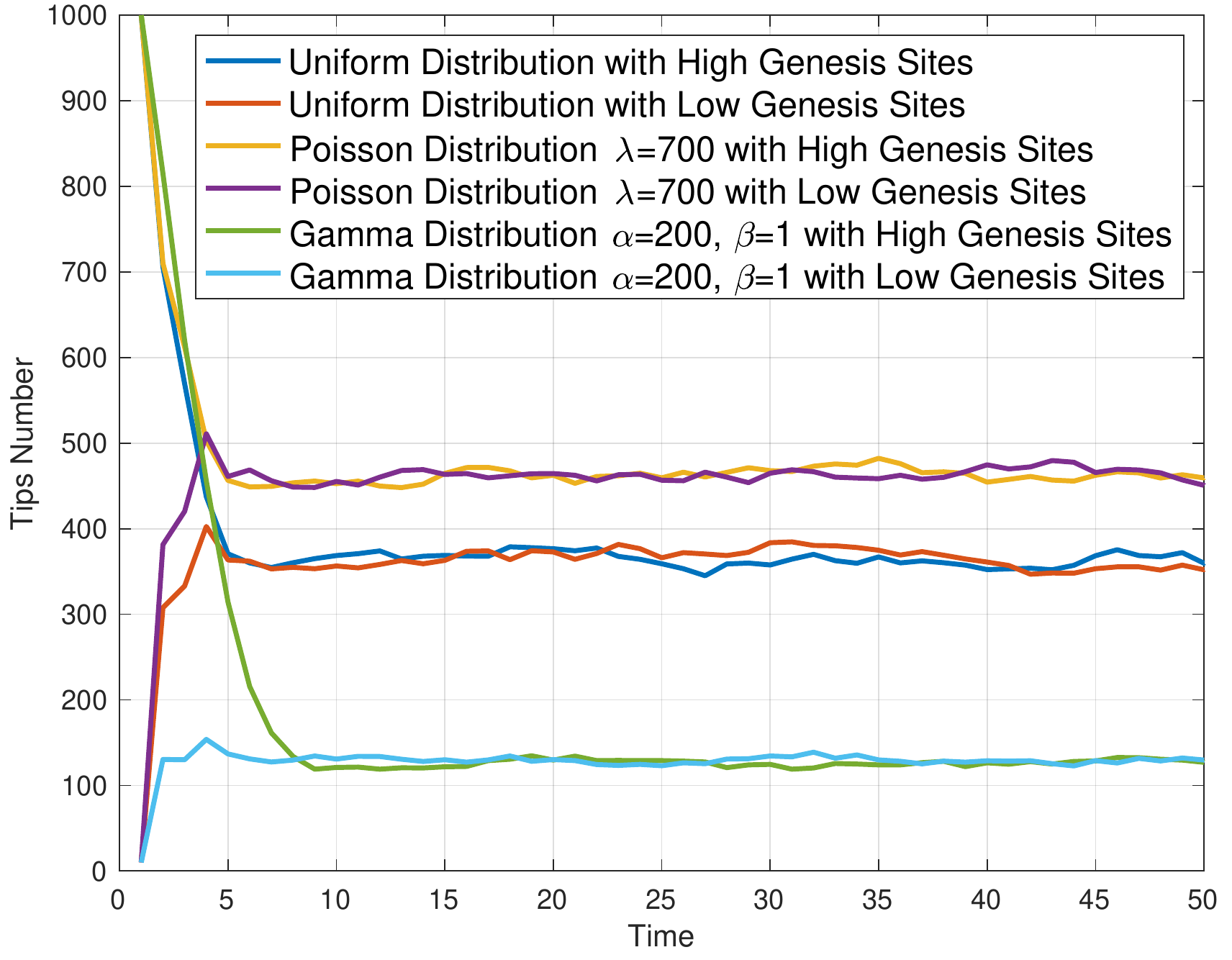}
\caption{Ledger Convergence under Different Tip-Arrival Distribution}
\end{figure}
\begin{figure}[t]
\centering
\includegraphics[scale=0.4]{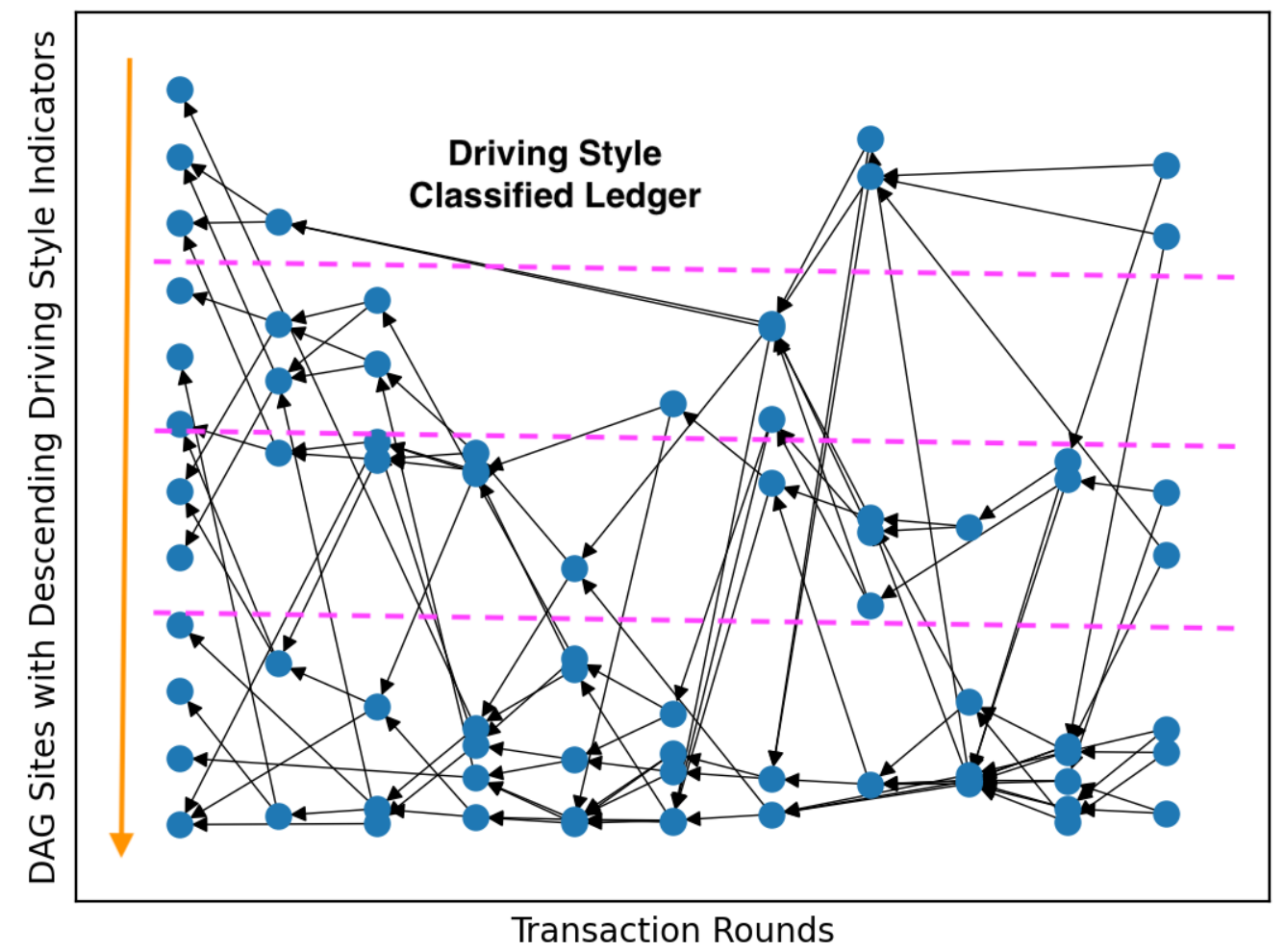}
\caption{Driving-Style Classified DAG Ledger}
\end{figure}

\begin{figure}[t]
\centering
\includegraphics[scale=0.5]{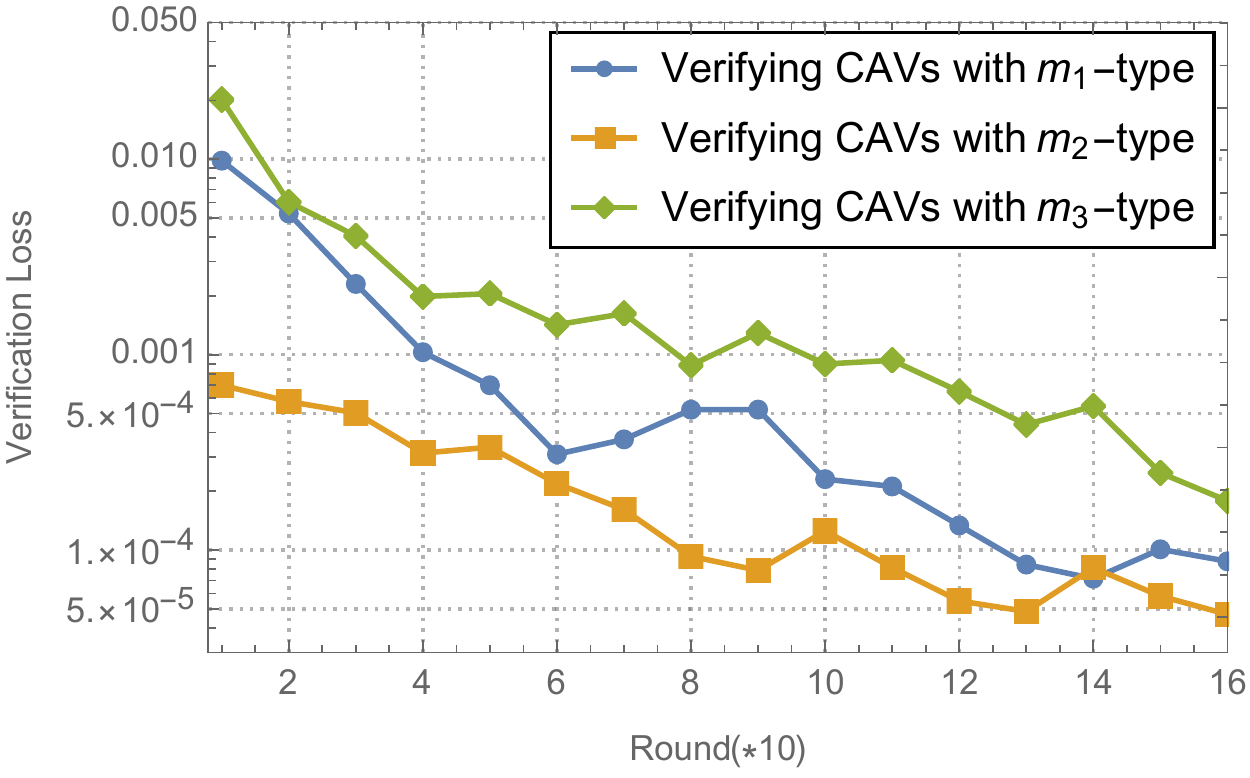}
\caption{Verification Loss of Different ICVs}
\end{figure}

\begin{figure}[t]
\centering
\includegraphics[scale=0.5]{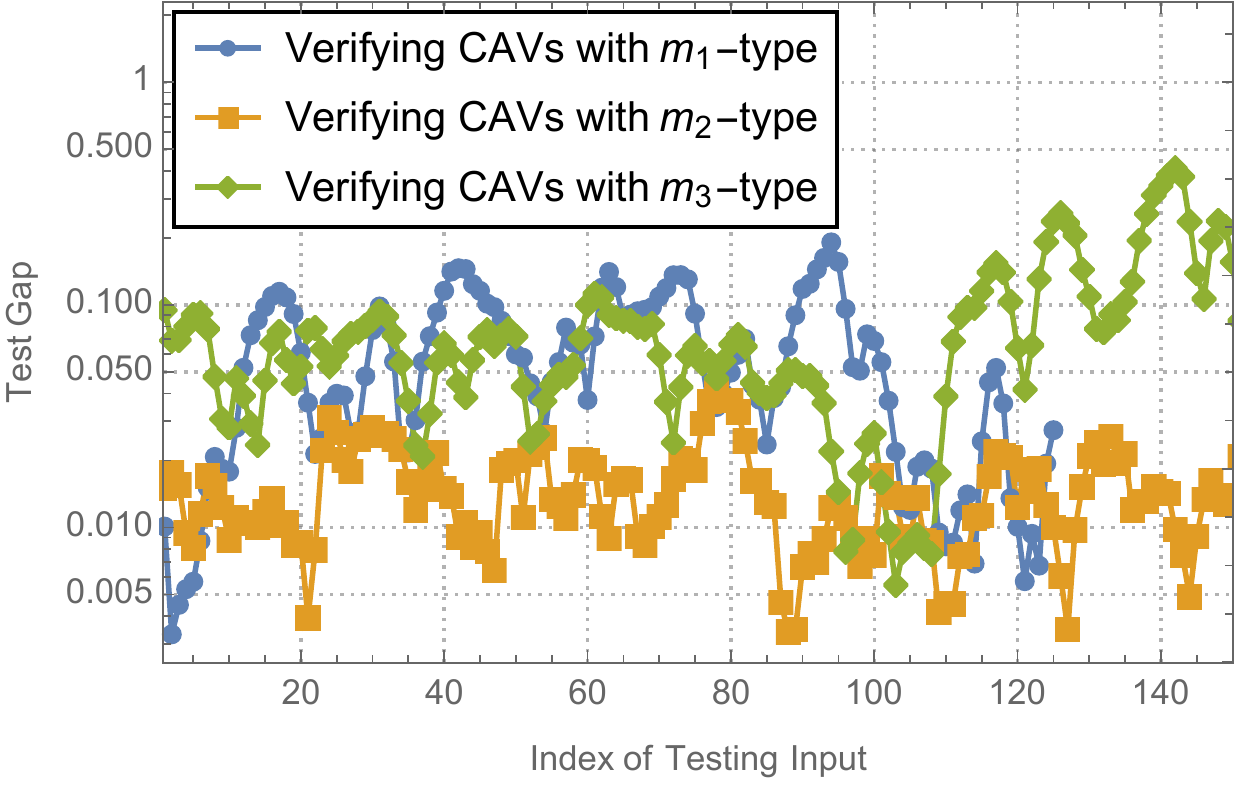}
\caption{Test Gap of the Shared Knowledge}
\end{figure}

The proposed driving style aware DAG framework is evaluated first.
The convergence results of the proposed DC-Ledger are shown in Fig. 5. Three tip-arrival distributions are considered, as indicated in Section. VI. A.
Two initial number of sites are set, $i.e.$ high genesis sites=1000 and low genesis sites=10. 
It can be found that no matter what the genesis number is, the final tip number will become stable, which shows the convergence of the proposed DAG ledger.

Fig. 6 presents the changes of DC-DAG ledger with different driving style for the genesis sites. 
The $y$-coordinate of the point in the figure represents the value of the driving style indicator.  
The driving style indicator value of the bottom point is close to 0, and the indicator value of the top point is close to 1.
We set the value of the driving style indicators of genesis sites ($i.e.$ points in the leftmost column) in a descending order, 
then we input incoming sites with different driving style indicators in each transaction round, which choose tips according to RTH-TSA in Eq. (4).
It can be obtained that with the transaction rounds, the total DAG ledger is logically divided into multiple sub-ledgers. 
This can be explained by the designed RTH-TSA:  According to Eq. (4), the new tips will firstly be appended behind the sites with similar driving styles.
With the tip selection algorithm processing, the ledger will eventually evolve to multiple driving style  classified sub-ledgers, 
each of which contains the sites with similar driving styles. 
The DC-ledger verifies the effectiveness of the proposed RTH-TSA, and it is beneficial for the site verification process.

\begin{figure*}
  \centering
  \subfigure[Loss Value of Global Model]
 {\includegraphics[width=2.3in,height=1.6in]{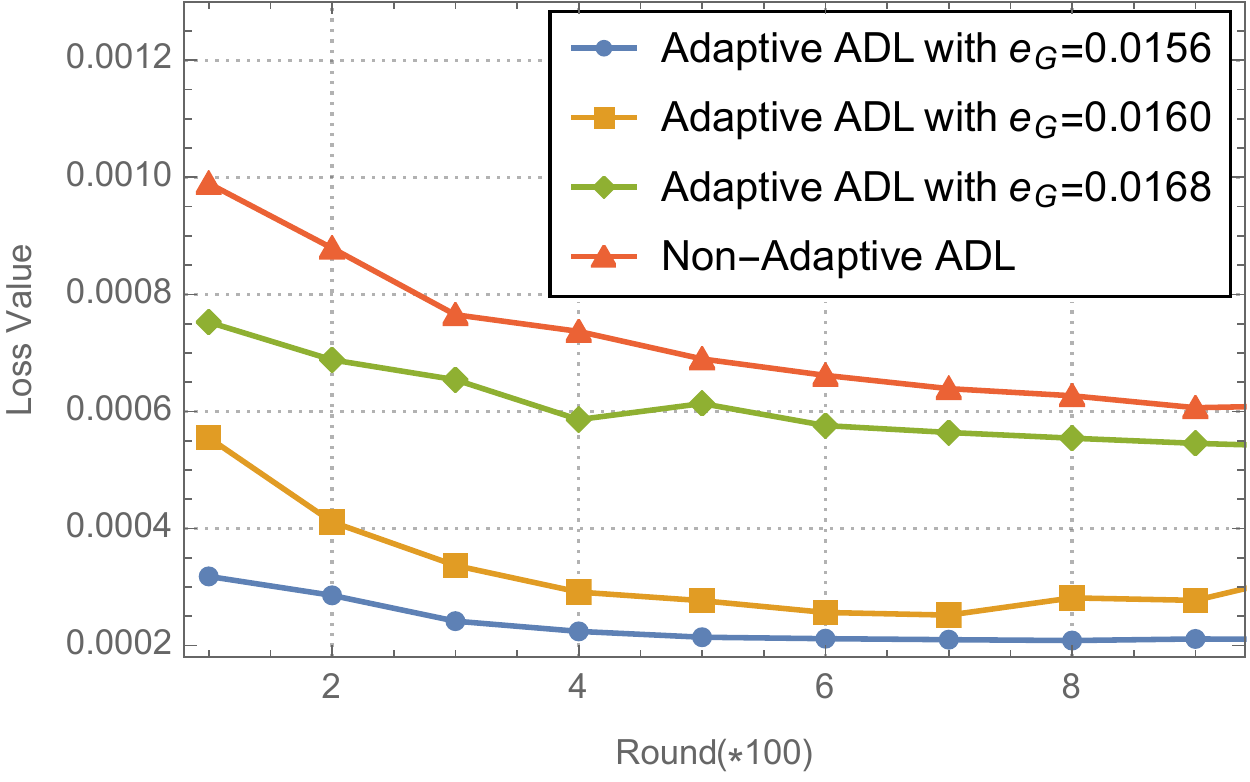}}
  \subfigure[Tset Gap]{\includegraphics[width=2.3in,height=1.6in]{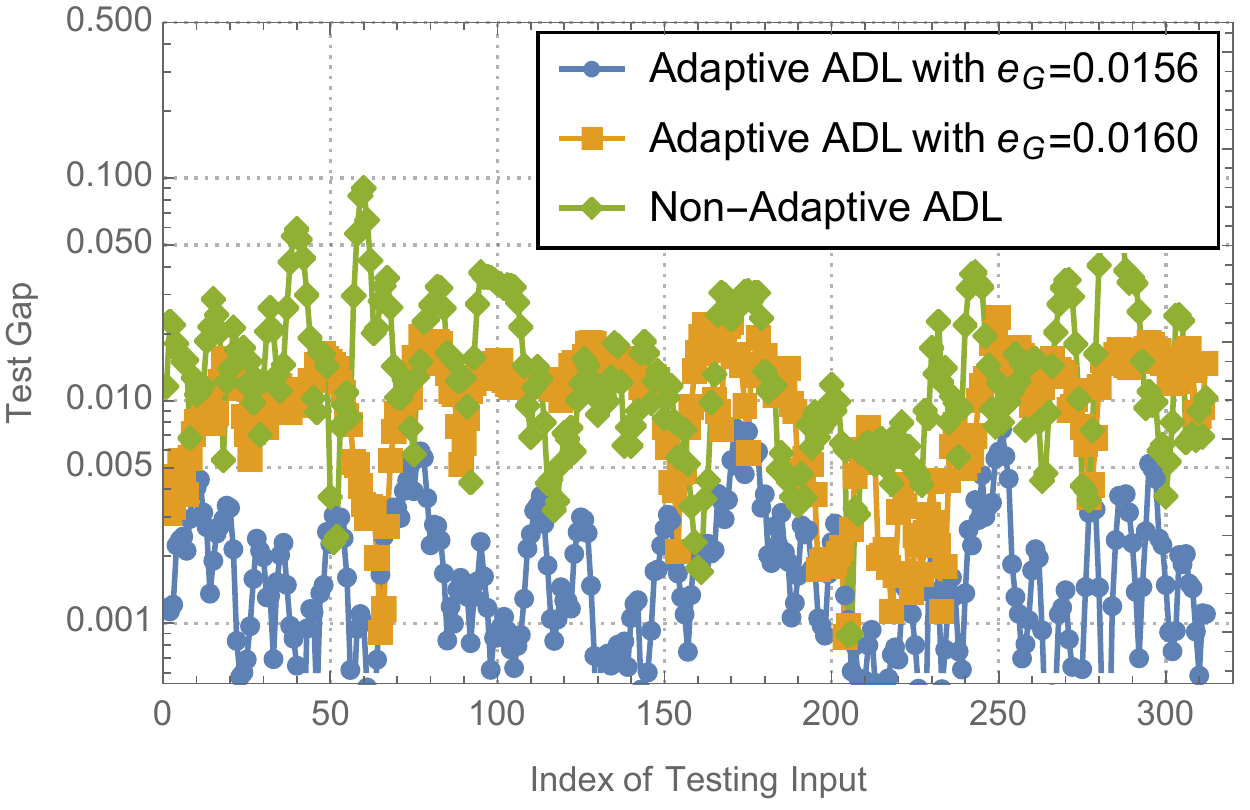}}
  \subfigure[Bandwidth Consumption]
{\includegraphics[width=2.3in]{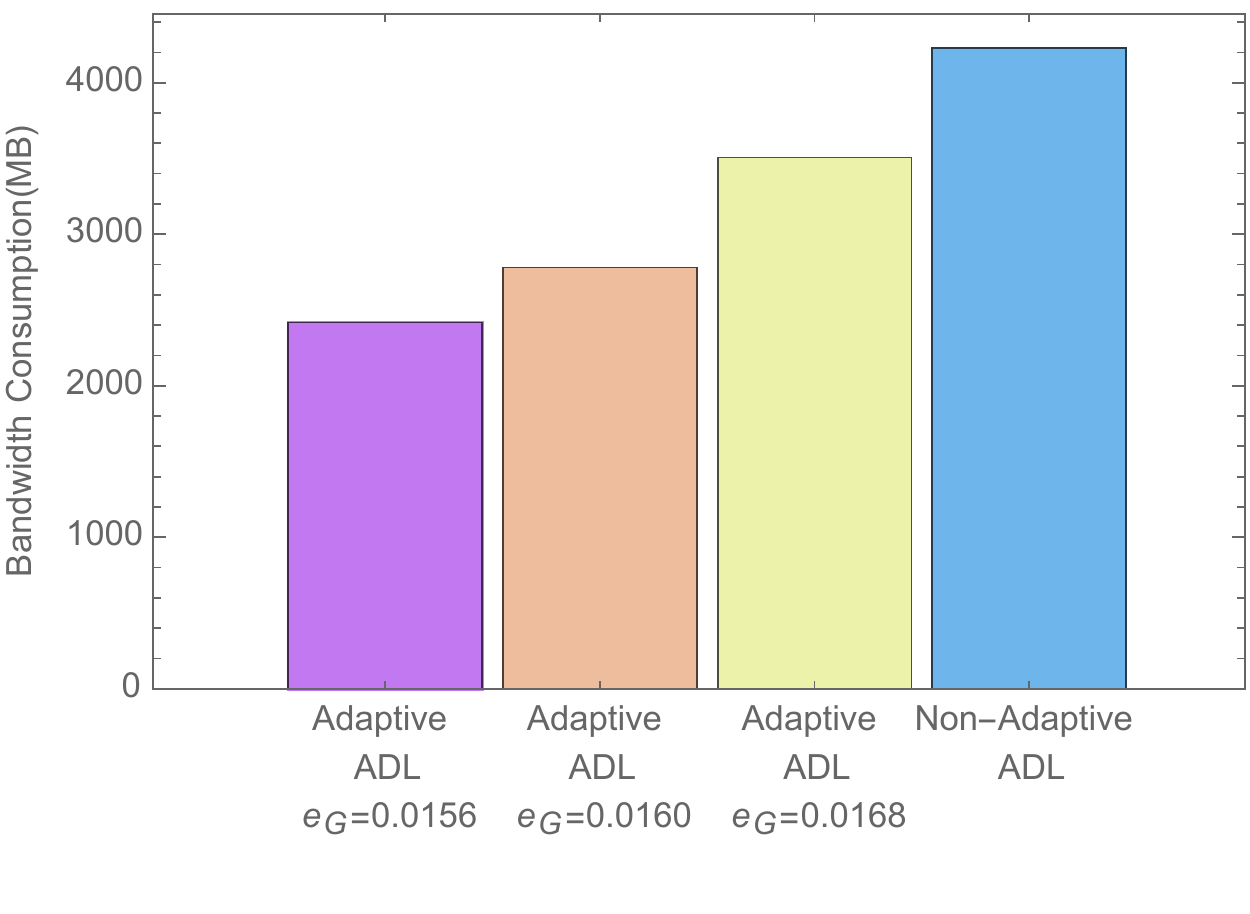}}
  \caption{Performance of the Adaptive ADL }
\end{figure*}

Based on the DC DAG ledger,  we evaluated the knowledge sharing efficiency in terms of verification loss and test gap.  
The verification loss is the loss function value produced by the verification of a pre-trained model over verification data. 
We let the three driving style types $m_1, m_2, m_3$ of ICVs to verify the adM that is trained by $m_2$-type ICVs.
The testing results are presented in Fig. 7 and Fig. 8.
In Fig. 7, it can be found that the ICVs with $m_2$-type have the minimal verification loss value.  
This indicates the verification loss can reach the optimal if the driving style indicator of ICVs matches that of the adM.
Note that the proposed RTH-TSA will enforce the sites to verify those sites with similar driving styles, and the verification loss value can be reduced.
That is revealed by the test gap in Fig. 8, where the three types of ICVs input their testing images and driving style indicators to the adM, 
and obtain the corresponding output control.  
As illustrated in the figure, the gaps of ICVs with $m_2$-type are less than ICVs with other types, 
which demonstrates the effectiveness of the proposed DAG based knowledge sharing framework. 
It also shows the potentials for future autonomous driving applications, 
where ICVs can exploit the model stored in the DC DAG to produce accurate autonomous driving control.

\begin{figure}[t]
\centering
\includegraphics[scale=0.43]{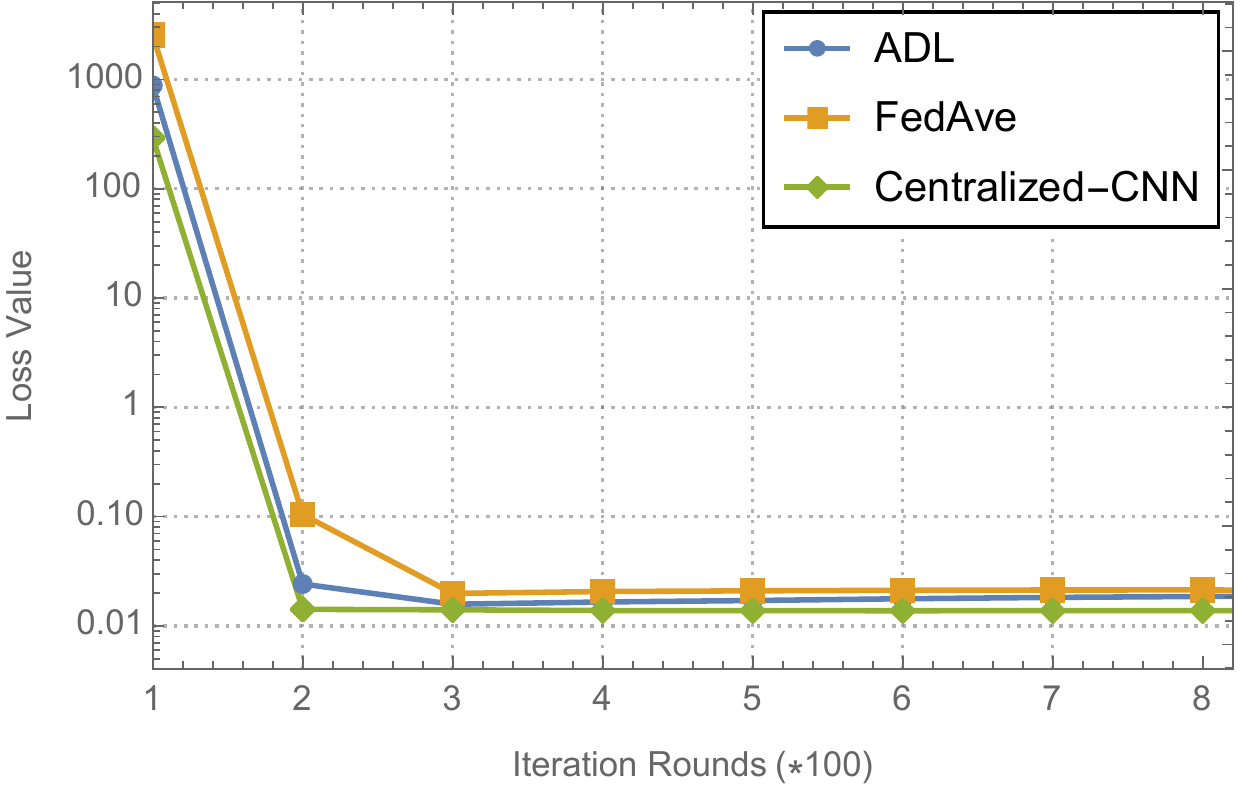}
\caption{Impact of Freshness on Learning Loss}
\end{figure}

Next, the proposed ADL is evaluated. We first investigate the proposed adaptive ADL scheme. 
The results of the adaptive ADL scheme are presented in Fig, 9 in terms of loss function, test gap and bandwidth consumption. 
We choose three different reference gaps $e_G=0.0156, 0.0160$ and $0.0168$. 
As can be seen in Fig. 9 (a), all the three adaptive ADL schemes outperform the non-adaptive ADL ones. 
This is due to that the reference gap $e_G$ filters some ICVs with poor local models.  
Only those ICVs with a local test gap $e_v > e_G$ are allowed to upload their adMs to the RSUs, 
which will improve the quality of global model. 
Besides, with the decrease of f $e_G$, the global loss value will be further reduced: a lower $e_G$ means that the RSUs have a better global model GadM.
Thus, in order to upload local adMs, the ICVs should also train a better CNN model, 
which will promote the entire network to evolve in a better direction. 
It is also noted that when the reference gap $e_G$ is updated to a sufficiently small value, there may be no more ICVs to upload their adMs. 
In this case, the global model can be deemed as converged and optimal.  
In terms of test gap in Fig. 9 (b), there is a similar trend to (a): compared to non-adaptive scheme that enforces all ICVs to upload their adMs,
the adaptive based schemes have a smaller gap. 
As the output of our CNN model is the steering value of ICVs, which is one of the crucial parameters for autonomous driving, 
by utilizing the adaptive based ADL, the RSUs can obtain a more accurate global model to implement autonomous driving.

The results of communication cost measured by the bandwidth consumption are presented in Fig. 9 (c), 
in which the bandwidth consumption refers to the volume of uploading adMs. 
The CNN model is provided by the Mathematica platform, which is consisted of two convolution layers,  two pooling layers, one linear layer, one flatten layer and three ramp layer.
The total network size is 120.9 MB.
Since the adaptive ADL scheme utilizes the reference gap $e_G$ to control the number of updated adMs, the consumed bandwidth will be reduced. 
As shown in the figure, the adaptive ADL with $e_G$=0.0156 can achieve a more than 30 \% reduction towards the non-adaptive scheme. 
By reducing the communication cost, the final leaning quality of the GadM does not degrade, which proves that the proposed scheme is suitable for the large-scale ICV networks.

\begin{figure}[t]
\centering
\includegraphics[scale=0.43]{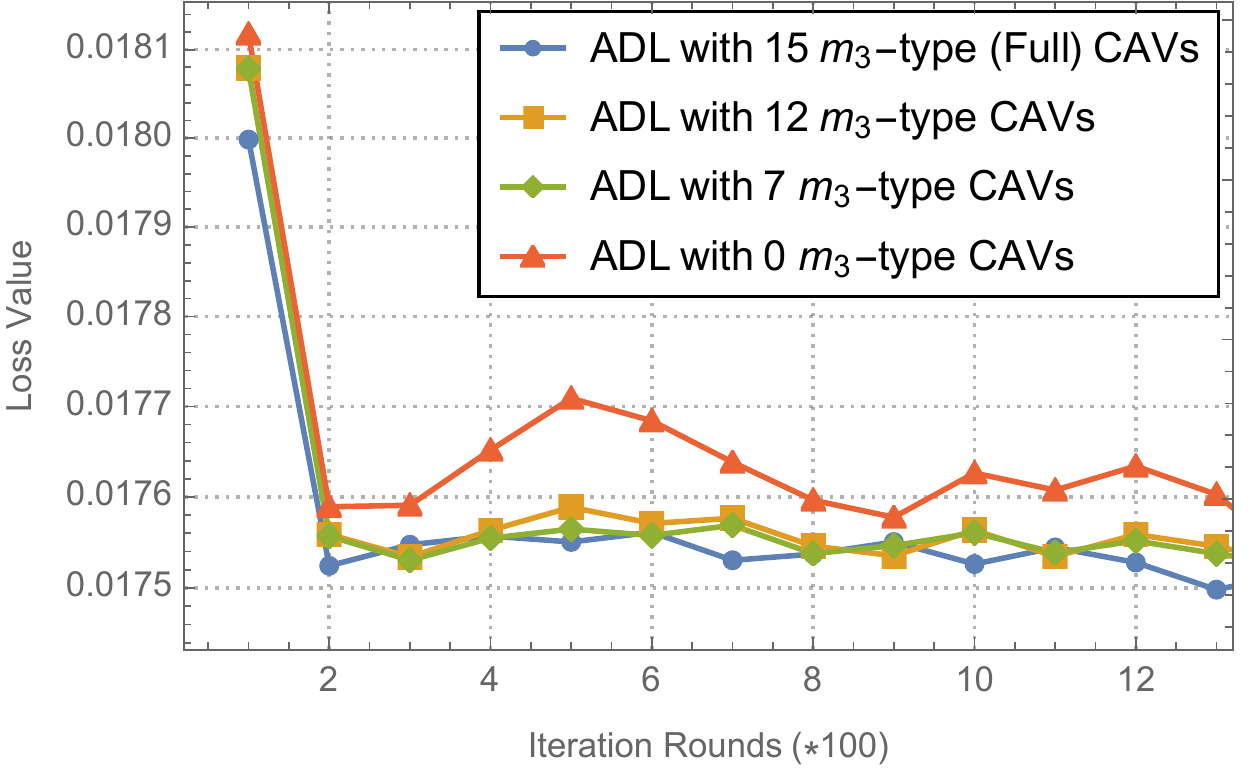}
\caption{Impact of Driving Style Indicator on Learning Loss}
\end{figure}

\begin{figure}[t]
\centering
\includegraphics[scale=0.43]{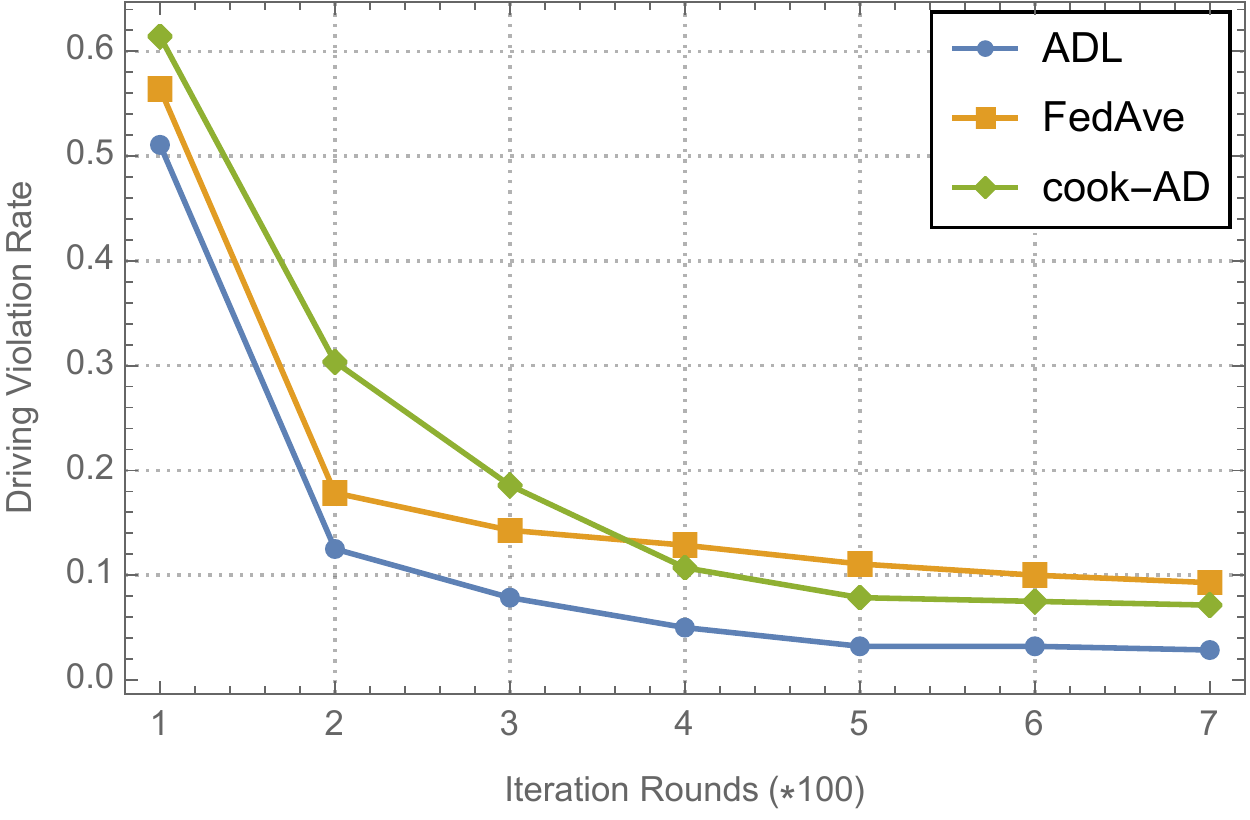}
\caption{Driving Violation Rate of different Mobility Styles}
\end{figure}

\begin{figure}[t]
\centering
\includegraphics[scale=0.43]{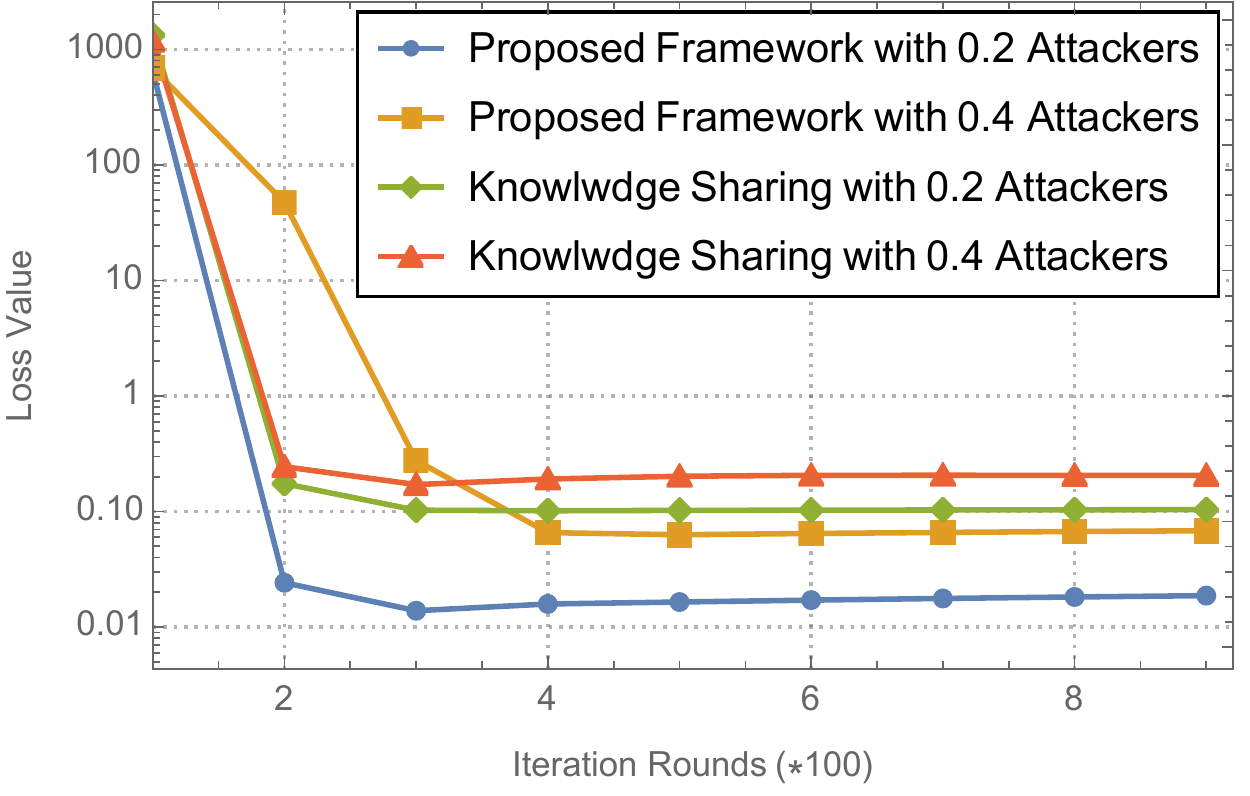}
\caption{The Security Performance under Attackers}
\end{figure}

Then, we investigate the impact of  freshness on the model learning loss. The results are presented in Fig. 10. 
The performance of the CNN model trained in a centralized approach is adopted here as the optimal training result. 
It can be seen that the proposed ADL has a loss very close to the optimal one, while that of FedAve is much higher. 
Therefore, from the perspective of global model training process, the subsequent local model tends to have a greater influence than previous models, as expressed in \emph{Remark 1}. 
The freshness ensures that the weight of subsequent models is higher than the earlier ones.

The results on the impact of driving style indicator on the model quality are shown in Fig. 11.  
The participant ICVs have test gap $e_v \leq e_G$,  which are capable of uploading their adMs.  
Here the total 35 vehicles are divided into three types $m_1$, $m_2$ and $m_3$, as shown in Table 1. 
The specific numbers of the ICVs with these three types are set as $N_{m_1+m_2}=20$, $N_{m_3}=15$.
And we set the average $\overline{M}=0.5$.  Since the driving style indicator of $m_3$-type is [0.095, 0.23], 
the $m_3$-type vehicles can be regraded as those ICVs with indicators far away from the average $\overline{M}$. 

As shown in Fig. 11, the loss values of both 12 ICVs with $m_3$-type and 7 ICVs with $m_3$-type are very close to that of full participants scenario. 
By reduce the number of those "deviated" ICVs,  we can reduce the communication cost of model updating,meanwhile ensuring the learning quality of global model, 
which verifies the claim in \emph{Remark 2}.  
Moreover, if we keep decrease the number of ICVs to the 0 $m_3$-type scenario in the figure, the learning loss will increase. 
This can be explained that although the deviated ICVs have little impact on the global model, their models do contain some information that other ICVs don't have. 
In order to obtain an accurate model, the final comprehensive model should aggregate some adMs from those deviated ICVs. 


The violation rate results are presented in Fig. 12. 
With the iteration process, the proposed ADL algorithm achieves a 20\% reduction in violation rate over the cook-AD and FedAve algorithms, 
which demonstrates the effectiveness of the proposed algorithm.  
It can also be found that both the proposed ADL and the FedAve algorithms converge faster than the cook-AD.
It can be explained by that utilizing the updating schemes, the distributed learning-based algorithms can collect local models from multiple distributed ICVs.
The distributed learning algorithm can accelerate the learning of the global context. 
Although the cook-AD algorithm utilizes a preprocessing method of flipping images that enlarges the training set, 
it will increase the complexity of training data and reduce the convergent speed.

At last the security of the proposed knowledge sharing framework is investigated with results shown in Fig. 13. 
Two scenarios with different proportions of attackers are considered with 0.2 and 0.4 attackers, respectively. 
The attackers attempt to share malicious knowledge with biased driving style indicators, and affect the final learned model.  
It can be found from the figure that the proposed framework achieves a lower loss value than that of the original knowledge sharing system.
The proposed one reduces the loss function value by an average of 30\%, which is mainly due to the use of DAG blockchain. 
The proposed DAG blockchain system utilizes the verification process $e \leq \epsilon$ to judge if the knowledge is legal, which can prevent the malicious data of attackers from being appended to the ledger.

\section{Conclusion}
In this paper, we proposed a DAG based knowledge sharing framework for ICVs to provide security and reliability under dynamic and mobile vehicle networks.
Built on the top of the framework, a specific application of knowledge sharing to autonomous driving control was investigated, 
wherein driving control related models were jointly trained and shared by ICVs.  
A RTH-TSA was proposed for tip selection in DAG to achieve lightweight consensus and fast cross-region identity authentication. 
Furthermore, we proposed an adaptive ADL scheme to enhance quality of shared knowledge as well as reducing communication cost during the learning process.
Experiment results  demonstrated the efficiency and stability of the proposed knowledge sharing framework. 
The proposed DAG based framework reduced a 30\% loss value against malicious attacks.
And the proposed adaptive ADL algorithm is effective and efficient, with more than 20\% reduction in terms of driving violation rate compared with the investigated existing algorithms.

\ifCLASSOPTIONcaptionsoff
  \newpage
\fi

\end{spacing}
\end{document}